\documentclass[
 reprint,
superscriptaddress,
 amsmath,amssymb,
 aps,
 prc,
]{revtex4-2}

\usepackage{physics}
\usepackage{amssymb}
\usepackage{amssymb}
\usepackage{amsfonts}
\usepackage{bbm}
\usepackage[colorlinks=true, allcolors=blue]{hyperref}
\usepackage{isotope}
\usepackage{tikz}
\usepackage{listings}
\usepackage{mathtools}
\usepackage{cleveref}
\usepackage{comment}
\crefname{section}{Sec.}{Secs.}
\Crefname{section}{Section}{Sections}

\newcommand{\sixj}[6]{\begin{Bmatrix}
#1&#3&#5\\
#2&#4&#6\end{Bmatrix}}
\newcommand {\IR} [1]{\textcolor{red}{#1}}

\newcommand {\IB} [1]{\textcolor{blue}{#1}}
\newcommand{\rme}[3]{\left\langle 
#1\left|\left|#2\right|\right|#3\right\rangle}
\newcommand {\ve} [1] {\mbox{\boldmath $#1$}}

\begin{document}

\title{Development of an accurate formalism to predict properties of two-neutron halo nuclei: case study of $^{22}$C}

\author{Patrick McGlynn}
\email[]{mcglynn@frib.msu.edu}
\affiliation{Facility for Rare Isotope Beams, Michigan State University, East Lansing, Michigan 48824 USA}
\author{Chlo\"e Hebborn}
\email[]{hebborn@ijclab.in2p3.fr}
\affiliation{Facility for Rare Isotope Beams, Michigan State University, East Lansing, Michigan 48824 USA}
\affiliation{Université Paris-Saclay, CNRS/IN2P3, IJCLab, 91405 Orsay, France}
\affiliation{Department of Physics and Astronomy, Michigan State University, East Lansing, Michigan 48824 USA}

\date{\today}
\begin{abstract}
\textbf{Background:} When moving away from stability or in loosely-bound systems, few-body clusterized structures such as two-neutron halo nuclei have been observed.  
These structures emerge due to the interplay between the many-body  and few-body degrees of freedom, and/or strong coupling between bound and continuum states. This motivates the development of models that can accurately describe few-body dynamics while enforcing shell effects.

\textbf{Purpose:} This work has two main goals: understanding how to accurately enforce the Pauli principle in few-body models, as well as presenting new technical developments that allow for more robust and less computationally-expensive three-body calculations. This study focuses on properties of the two-neutron halo $^{22}$C, but we expect the conclusions to apply to other few-body systems.

\textbf{Methods:} We consider a three-body model using the hyperspherical harmonics formalism combined with the R-matrix method. We compare predictions for properties of $^{22}$C, starting from phenomenological interactions, and using two methods to remove Pauli-forbidden states, the projection and supersymmetric methods.  We also present the algorithms and derivations used for the implementation of this formalism. Additionally, we explore truncations of the model space that allow for reduced computational time.

\textbf{Results:} We show convergence of the calculation of both bound and scattering states for $K_{max}\sim$~$40$. The two methods to enforce the Pauli-exclusion principle lead to different predictions of $^{22}$C properties. Our detailed study shows that the projection method is more accurate.  We find one efficient truncation in the number of channels considered that reduces the computational cost of our calculations by $20$\%.

\textbf{Conclusions:} Our in-depth study clarifies that the projection method is more accurate than the supersymmetric one to enforce the Pauli-exclusion principle. We also demonstrate that technical and algorithmic developments  enable us to compute accurately and efficiently properties of two-neutron halo nuclei. This development paves the way to robust uncertainty quantification in three-body predictions, and constitutes a useful starting point to tackle more complex systems and observables.
\end{abstract}

\maketitle

\section{Introduction}

During the last decades, the emergence of radioactive ion beams  (RIB) has enabled the study of unstable nuclei at the edge of the nuclear chart and the discovery of new exotic phenomena. In particular,  clusterised structures were found  in loosely-bound systems, located close to the driplines and in states below the threshold (see Refs.~\cite{FewbodyPerspectives,Freer2018} and references therein).  The most archetypical examples being halo nuclei~\cite{TANIHATA2013215},  in which the outermost bound nucleons exploit quantum tunnelling to extend their wavefunctions far outside the core. Two-neutron halo nuclei are even more exotic as they exhibit a Borromean character, in which the three-body system is bound, while each of its constituent two-body subsystems is unbound.  It is understood that the emergence of such clusterisation results from a complex interplay of few-body dynamics,  coupling to unbound states, deformation and shell effects~\cite{FewbodyPerspectives,11BeNCSMC,PhysRevC.106.034312,PhysRevLett.128.212501,HammerHalo,Orr2003}.  Even though halo nuclei still represent  a challenge for most theoretical models, accurately predicting their properties is still an important goal as it will provide a deeper understanding of the transition from many-body degrees of freedom to few-body effects in nuclear physics, and in general in quantum many-body systems. 

Experimentally, several one- and two-neutron halo systems were measured and studied at RIB facilities (see for example Refs.~\cite{PhysRevLett.122.052501,PhysRevLett.108.192701,29FNature,8BNature,Cook2020,Aumann1999,Wang2002,Sackett1993,Ieki1993,Zinser1997,Nakamura2006,Nakamura2017}). Most studies focus on studying the  energy of their states, their spin and parities, their root-mean-square (r.m.s) radii, providing information about the spatial extension of the nucleus, as well as the dipole strength, known to be large as the nuclear center-of-mass and center-of-charge do not coincide (see review in Ref.~\cite{AumannNakamuraReview}). In many instances, it has been observed that the emergence of  halo structures is accompanied by unexpected shell structure, such as  shell inversion~\cite{29FNature,11Bedata,11Bedata2}. Theoretically, models starting from nucleonic degrees-of-freedom tend to struggle  to reproduce these halo nuclei, since they often rely on an expansion of the wavefunction onto harmonic oscillator bases which limit the spatial extension of the wavefunction and effectively discretize the continuum states. Few-body models, seeing the nucleus as a core and halo nucleons, treat these few-body effects by construction, but at the price that the shell structure does not emerge from the calculations, and is enforced a priori.

Typically,  information about shell structure is included in few-body models in two steps (for examples see Refs. \cite{Baye2009,Pinilla2016,Casal2020,Tostevin2001,Nunes1996a,Nunes1996b}). First by tuning the effective interaction between the clusters to reproduce experimental information about their shell structure, e.g. by adjusting a resonance position. Second, by enforcing the Pauli principle between the nucleons through the removal of Pauli-forbidden states. This is done through supersymmetric transformation of the two body potentials~\cite{Baye1987} or by explicitly projecting out these states~\cite{Kukulin1978}. Although both methods are commonly used,  previous works focusing on $^{6}$He~\cite{Thompson2000,Descouvemont2003} have shown that the choice of the method influences the properties of two-neutron halo nuclei. Nevertheless, the comparison of both methods has only been done for light systems, $^{6}$He~\cite{Thompson2000,Descouvemont2003} and $^6$Li~\cite{TURSUNOV2020121884}, which contain only one forbidden states per two-body subsystem. 

The first part of our work focuses on further investigating the impact of the choice of the method to remove Pauli-forbidden states. To this end, we use as in Refs.~\cite{Descouvemont2003,Descouvemont2006} the hyperspherical harmonics method coupled with the R-matrix method~\cite{Descouvemont2010}. The main advantage of this approach is that  it treats consistently bound and scattering states and does not discretize the continuum. We perform a detailed comparison between both methods to remove Pauli-forbidden states studying the properties of $^{22}$C. Compared to $^6$He, $^{22}$C  contains more than one Pauli-forbidden state to be removed and has its valence neutrons in the $sd$ region. Our detailed study  enables us to provide recommendations on which method is more reliable.

Because it is expected that  other halo structures in the mid-mass region will be discovered with the new capabilities of RIB facilities such as FRIB, it is also pressing to develop accurate codes. Moreover, to make meaningful comparisons with these future experimental data, the theoretical uncertainties need to be quantified. This often relies on running our code multiple times to gather enough statistics, motivating the development of an efficient code making extensive use of modern computing architecture. This work also presents the theory and implementation of our new code {\sc hyperboromir}, as well as  various technical developments that allow for a simpler and more efficient implementation of this formalism. In particular, we develop new algorithms to apply the projection method to remove Pauli-forbidden states and to automate of the adjustment of a three-body force. We also derive a general expression for $B(E\lambda)$ that can be used in few-body models to compute general electric mulitpole strengths.  We finally study new truncations of the model spaces that allow for reducing the computational time, without reducing the accuracy of the three-body predictions.

The paper is structured as follows. In Section~\ref{Sec:3b}, we introduce the basis we use to solve the three-body problem and  both methods used to remove Pauli forbidden states. In  Section~\ref{sec:Rmatrix}, we show how bound and scattering states are calculated in the R-matrix method.  Section~\ref{sec:Be1} shows the form of the electric dipole strength function in this basis. In Section~\ref{sec:potentials} we explain the specific choice of $^{22}$C as a three-body system and specify the potential(s) used to represent both core-neutron, neutron-neutron interactions and three-body interactions.  Section~\ref{sec:resultsSusyProjection} compares the $^{22}$C properties obtained with both the supersymmetric and projection method. Section~\ref{Sec:Speedup}  discusses the computational costs of the method and presents new avenues to decrease it. Finally, Section~\ref{Sec:Conclusions} presents the conclusions and prospects of this work. Because our work  contains several technical developments, we make also an extensive use of appendices to detail them.

\section{Three-body formalism} \label{Sec:3b}
\subsection{Hyperspherical harmonics}
\label{sec:hyperspherical}

In this work, we are interested in describing systems exhibiting a clusterized three-body structure. Although our study will focus on two-neutron halo nuclei, we will  present the hyperspherical formalism for a general three-body system, composed of three clusters with spins $I_1$, $I_2$ and $I_3$ and mass numbers $A_1$, $A_2$ and $A_3$ (more details on this formalism can also be found in Refs.~\cite{Descouvemont2003,Descouvemont2006,MarcucciHH}). The Hamiltonian of this three-body system reads
\begin{equation}
    H=\sum_{i}T_{\ve r_i}+\sum_{i<j}V_{ij}(\mathbf{r}_i-\mathbf{r}_j)+V_{3b}(\mathbf{r}_1,\mathbf{r}_2,\mathbf{r}_3),
\end{equation}
where $\ve r_i$ are the single-particle coordinates of each cluster, $T_{\ve r_i}$ are the corresponding  kinetic energies, $V_{ij}$ and $V_{3b}$ are respectively the  two- and three-body forces. 

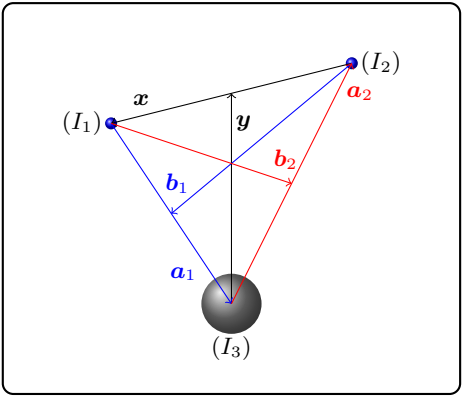
\begin{figure}

 \centering
    \begin{tikzpicture}[scale=0.8,every text node part/.style={align=center}]
    \shade[ball color=gray] (0,-3) circle (0.5cm);
    
    \shade[ball color=blue] (-2,0) circle (0.1cm);
    \shade[ball color=blue] (2,1) circle (0.1cm);


    \node[below] at (0,-3.4) {($I_3$)};
    \node[left] at (-2,0) {($I_1$)};
    \node[right] at (2,1) {($I_2$)};
    \draw[<-] (-2,0) -- (2,1); 
  \draw (-1.5,0.4)    node {$\ve x$};
    \draw[<-] (0,0.5) -- (0,-3); 
      \draw (0.2,-0)    node {$\ve y$};
    \draw[->,blue] (-2,0) -- (0,-3); 
            \draw (-0.8,-2.5) node{\IB{$\ve a_1$}};

    \draw[<-,blue] (-1,-1.5) --(2,1); 
    \draw (-0.9,-1) node{\IB{$\ve b_1$}};
      \draw[<-,red] (2,1) -- (0,-3); 
   \draw (2.2,0.5)    node {\IR{$\ve a_2$} };
    \draw[->,red]  (-2,0)-- (1,-1); 
        \draw (0.9,-0.6) node{\IR{$\ve b_2$}};

    \draw[thick,rounded corners] (-3.8,-4.5) rectangle (3.8,2);

\end{tikzpicture}
\caption{Definition of T-basis (in black)and Y-basis (in red and blue)  coordinates and particle spins \label{fig:TandYbasisdiagram}.}
\end{figure}

As usual in few-body frameworks, we work in Jacobi coordinates (Fig.~\ref{fig:TandYbasisdiagram}) to decouple the centre-of-mass motion. The Jacobi coordinates in the T-basis are defined as
\begin{align}
    \ve x&=\sqrt{\mu_{12}}(\ve r_1-\ve r_2)\\
   \ve  y&=\sqrt{\mu_{(12)3}}\left(\ve r_3-\frac{A_1\ve r_1+A_2\ve r_2}{A_1+A_2}\right),
\end{align}
where 
  \begin{align} 
    \mu_{12}&=\frac{A_1A_2}{A_1+A_2},\\
    \mu_{(12)3}&=\frac{(A_1+A_2)A_3}{A_1+A_2+A_3}
\end{align}
 are the unitless reduced masses, where fragment has mass $m_i=A_im_N$ with $m_N$ the nucleon mass. Equivalent  coordinates in the Y-bases  (which we call $(\ve a_1,\ve b_1)$ and $(\ve a_2, \ve b_2)$) can be obtained  by permuting the indices.

The hyperspherical coordinates are  the solid angle $\Omega_{x,y}$ associated with  $\ve x$ and $\ve y$, the hyperradius $\rho$ and the hyperangle $\alpha$, defined from the Jacobi coordinates as
\begin{align}
    \rho^2&=|x|^2+|y|^2\\
    \alpha&=\arctan\frac{|y|}{|x|}
\end{align}
Using hyperspherical coordinates, one can then write the Hamiltonian of the three-body system, free of the center-of-mass motion, 
\begin{equation}
    \mathcal{H}=T_\rho+\sum_{n<m}V_{nm}+V_{3b} \label{eq8}
\end{equation}
where $ T_\rho$ is the kinetic operator 
\begin{equation}
    T_\rho=-\frac{\hbar^2}{2m_N}\left(\pdv[2]{\rho}+\frac{5}{\rho}\pdv{\rho}-\frac{K^2}{\rho^2}\right)\label{eq9}
\end{equation}
with $K$ called the hyperangular momentum. In this formalism, one chooses to expand the wavefunction onto the eigenstates of the $K^2$ operator $\mathcal{Y}$, called hyperspherical harmonics, i.e. 
\begin{equation}
    \Psi^{JM\pi}(\rho,\Omega_{5\rho})=\rho^{-5/2}\sum_{K=0}^\infty\sum_\gamma\chi_{K \gamma }^{J\pi}(\rho)\mathcal{Y}_{K \gamma}^{JM}(\Omega_{5\rho}) \label{eq10}
\end{equation}
with~\footnote{The pre-factor of $\rho^{-5/2}$ in \cref{eq10} is introduced  to eliminate the first derivative term in \cref{eq9}.}
\begin{equation}
    \mathcal{Y}_{K\gamma}^{JM}=\phi_{K}^{l_xl_y}(\alpha)[[Y_{l_x}(\Omega_x),[\chi_{I_1},\chi_{I_2}]_s,Y_{l_y}(\Omega_y)]_L,\chi_{I_3}]_{JM}\label{eq:coupling}.
\end{equation}
$\Omega_5=(\alpha,\Omega_x,\Omega_y)$,  $Y_l(\Omega)$ are  the usual spherical  harmonics, $\chi$ are spinors, and the brackets denote coupling with sums over magnetic quantum number and the appropriate Clebsch-Gordan coefficients. The shorthand $\gamma$ refers to the quantum numbers $l_x,I_1,I_2,s,l_y,L,I_3$, with $l$ denoting orbital angular momenta. 
In practice, the sum over $K$  in \cref{eq10} is bounded by some finite $K_{max}$, which has to be taken large enough to reach convergence.

The hyperangular function $\phi_{K}^{l_xl_y}$ appearing in \cref{eq:coupling} read 
\begin{align}
    \phi_{K}^{l_xl_y}(\alpha)&=\mathcal{N}_K^{l_xl_y}(\cos\alpha)^{l_x}(\sin\alpha)^{l_y}P_n^{l_y+1/2,l_x+1/2}(\cos2\alpha),\\
  \text{with }  \mathcal{N}_K^{l_xl_y}&=\sqrt{\frac{2n!(K+2)(n+l_x+l_y+1)!}{\Gamma(n+l_x+3/2)\Gamma(n+l_y+3/2)}},\\
    n&=\frac{K-l_x-l_y}{2},
\end{align}
 and $P_n^{a,b}$ Jacobi polynomials. Since $n$ can only be a positive integer, we have $K\geq l_x+l_y$ and $K\pmod 2\equiv l_x+l_y\pmod 2$.
 
Inserting \cref{eq10} into \cref{eq8}, one can show that hyperradial wavefunctions $\chi_{K\gamma}^{J\pi}$ are solution to the hyperradial Schrödinger equation
    \begin{align}
    [\mathcal{T}\, &\delta_{K\gamma,K'\gamma'}+\sum_{K'\gamma'}V^{K\gamma}_ {K'\gamma'}(\rho)]\chi_{K' \gamma '}^{J\pi}(\rho)=E\chi_{K \gamma }^{J\pi}(\rho)\\
    \text{with } \mathcal{T}&=-\frac{\hbar^2}{2m_N}\left(\pdv[2]{\rho}-\frac{(K+3/2)(K+5/2)}{\rho^2}\right), \label{eq:schrodinger}
\end{align}
and $V^{K\gamma}_ {K'\gamma'}=\mel{\phi_K^{l_xl_y}}{V}{\phi_{K'}^{l_x'l_y'}}$  are the coupling potentials. These potentials are obtained by integrating over the hyperangular coordinate $\alpha$ 
to project the two- and three-body forces in \cref{eq8} onto the basis states in \cref{eq10}. 

In this work, we consider  two-neutron halo systems, for which two of the clusters, chosen here as (1) and (2), are neutrons $(n)$, and the last cluster (3) is the core  $(c)$.  To ensure  antisymmetry of the wavefunction under the exchange of the two neutrons, the impose the condition $(-1)^{l_x+s}=1$.  We evaluate the $n$-$n$ interaction $V_{12}$ in the T-basis and  the core-neutron ($cn$) interaction $V_{13}=V_{23}$ in the Y-bases. We also consider a three-body force $V_{3b}$ that does not couple different channels and thus can be evaluated easily in all bases. Ultimately, we calculate the total hyperradial potential in the T-basis, using the Raynal-Revai transformations \cite{Raynal1970,Youping1987} to transform  the matrix elements from one basis to another. For example, the $V_{13}$ matrix elements in the Y-basis, $(V_{13})^{K\gamma_Y}_{K'\gamma_Y'}$, can be expressed in the T-basis using
\begin{align}
(V&_{13})^{K\gamma_T}_{K'\gamma_T'}(\rho)=\nonumber\\
&\sum_{\gamma_Y\gamma_Y'}\mathcal{R}_K(\gamma_Y,\gamma_T)(V_{13})^{K\gamma_Y}_{K'\gamma_Y'}(\rho)\mathcal{R}_K(\gamma_Y',\gamma_T')\\
\text{with }\mathcal{R}&_K(\gamma_Y,\gamma_T)=\bra{\mathcal{Y}_{K\gamma_Y}}\ket{\mathcal{Y}_{K\gamma_T}}
\end{align}
More details  are given in appendix \ref{sec:RRtransform}.

\subsection{Removal of Pauli-forbidden states}

Because we are not considering the internal structure of each cluster, the Pauli principle is not enforced between nucleons inside the clusters. This translates into the presence of Pauli-forbidden two-body bound states, that lead to spurious three-body states. In two-neutron halo systems, those Pauli-forbidden   states are present in the core-neutron systems. 
In the literature, two methods have predominated to remove these forbidden  states from the three-body calculation: supersymmetric transformation~\cite{Baye1987} and a projection method~\cite{Kukulin1978}.

\subsubsection{Supersymmetric transformation}
The supersymmetric transformation modifies the two-body potential to eliminate a bound state, without changing the \textbf{two-body} phase shifts. The specific modification is an additional short-range repulsion of the form~\cite{Baye1987}
\begin{equation}
    V^{susy}_{lsj}(r)=-\frac{\hbar^2}{\mu_{cn}}\frac{d^2}{dr^2}\ln \left[\int_0^r dr' u_{nlsj}(r')^2\right]
\end{equation}
where $r$ is the radial distance between the two clusters, $u_{nlsj}(r)$ is the radial wave function associated with the bound wavefunctions whose quantum numbers   $nlsj$ are respectively the principal quantum number, orbital angular momentum, spin and total angular momentum. The supersymmetric potential is added in each partial wave where a bound state exists.

Computationally, the two-body bound states are calculated with a fine Lagrange mesh in order to obtain precise forms of $u_{nlsj}(r)$. 
The singularity at $r=0$ makes it necessary to calculate the wavefunction beginning from some $r>0$, but this leads to numerical difficulties in evaluating the supersymmetric potential at extremely small $r$. To resolve this, for $r<r_1$ (the first Lagrange point in the two-body mesh), we use the analytic form $V^{susy}(r)\propto r^{-2}$ which should apply for sufficiently small $r$, and match the proportionality constant to $V^{susy}(r_1)$.

As already seen in previous works~\cite{Capel2007,Thompson2000}, adding this short-range potential changes the two-body wavefunctions: the two-body asymptotic normalization constant and phase shifts are preserved but the solution to the transformed potential does not have the nodes corresponding to the removed states. 

\subsubsection{Projection method}
The projection technique is conceptually simpler but more computationally expensive. It involves computing the exact projection operator $\mathcal{P}$ which acts in the three-body Hilbert space to project only the forbidden states. Then, the three-body Hamiltonian is transformed to 
\begin{equation}
\mathcal{H}^{proj}=(1-\mathcal{P})\mathcal{H}(1-\mathcal{P})+\lambda\mathcal{P}
\end{equation}
where $\lambda$ is a very large real number ($10^{11}$ is chosen in this work) used to move the forbidden subspace from zero energy to an unphysically high one. The procedure for computing $\mathcal{P}$ is covered in Appendix~\ref{sec:projection}.

\section{Calculation of bound and scattering states} 
\label{sec:Rmatrix}
To solve \cref{eq:schrodinger} we apply the calculable $R$-matrix method. This method has the advantage of  treating consistently both bound and scattering states without discretizing the continuum. This is done by enforcing the appropriate boundary condition, and hence requires an analytic solution  for the asymptotic form of the  wavefunction.
 The formalism and its application to three-body problems has been discussed before in e.g. Refs.~\cite{Descouvemont2003,Descouvemont2006,Descouvemont2010} but key details are repeated here for  reference. 
  
 The essence of this approach is to solve the Bloch-Schr\"odinger equation  
\begin{align}
    [\mathcal{H}&+\mathcal{L}(B)-E]\chi^{\text{int}}=\mathcal{L}(B)\chi^{\text{ext}}\label{eq:blochschrodinger}\\
    \mathcal{L}(B)&=\frac{\hbar^2}{2m_N}\delta(\rho-\rho_{max})\left(\dv{\rho}-\frac{B}{\rho}\right) \label{eqBlochOp}
\end{align}
with $\mathcal{H}$ being a general notation for the Hamiltonian using either the supersymmetry or projection method to remove Paui-forbidden states. $\mathcal{L}(B)$ is the Bloch operator, which depends on the boundary-value parameter $B$,  $\chi^{\text{int}}$ is the wavefunction in the internal region ($\rho\leq\rho_{max}$) and $\chi^{\text{ext}}$ is the known external solution for $\rho\geq\rho_{max}$. The Bloch operator acts at the boundary and enforces continuity of the first derivative of the wavefunction. We solve \cref{eq:blochschrodinger} along with the boundary conditions $\chi^{\text{int}}(0)=0$ and $\chi^{\text{int}}(\rho_{max})=\chi^{\text{ext}}(\rho_{max})$. If $\rho_{max}$ is large enough that off-diagonal couplings in the external region are negligible, this method is exact.   

To solve the three-body Bloch-Schrödinger hyperradial equation \cref{eq:blochschrodinger}, we expand the  wavefunction  in the internal region onto a hyperradial mesh. The details of this mesh are provided in Appendix~\ref{AppHyperradialmesh}. 

\subsection{Bound states}

For bound states $(E<0)$, the known external solution~\cite{Descouvemont2006} is 
\begin{equation}
    \chi^{\text{ext}}_{K \gamma}(\rho)=\mathcal{C}_{K\gamma}(\kappa\rho)^{1/2}\mathcal{K}_{K+2}(\kappa\rho) \label{eq:boundext}
\end{equation}
where $\kappa=\sqrt{-2m_NE}/\hbar$, $\mathcal{K}_{n}$ is a modified Bessel function of the second kind, and $\mathcal{C}_{K\gamma}$ is  the three-body asymptotic normalisation coefficient (ANC) in the channel $K\gamma$. In order to solve \cref{eq:blochschrodinger} for negative energy we set $\mathcal{L}(B)\chi^{\text{ext}}=0$, by choosing the boundary value as 
\begin{equation}
    B=\frac{1}{2}+\kappa\rho_{max}\frac{\mathcal{K}'_{K+2}(\kappa\rho_{max})}{\mathcal{K}_{K+2}(\kappa\rho_{max})}\label{eq:Bvalue}
\end{equation}
so that the right hand side of \cref{eq:blochschrodinger} becomes 0. 

Since the value of $\kappa$ is not known beforehand, the bound states are found using an iterative algorithm. It first computes the eigen-pairs of  the $C$-matrix, defined as
\begin{align}
    &(C^{K\gamma }_{K'\gamma' })_{ij}=(\mathcal{T}^{K\gamma }_{K\gamma })_{ij}\delta_{KK'}\delta_{\gamma\gamma'}\nonumber \\
    &\quad+(\sum_{n<m}V_{nm}(\rho_i)+V_{3b}(\rho_i))^{K\gamma}_{K'\gamma'}\delta_{ij}+(\mathcal{L}^{K\gamma }_{K'\gamma' }(B))_{ij}.\label{eqCmatrix}
\end{align}
For negative eigenvalues, we determine their corresponding $\kappa$ and recompute the boundary condition \cref{eq:Bvalue}, the $C$-matrix~\cref{eqCmatrix} and rediagonalize the $C$-matrix. Convergence  is achieved when the eigenvalue of $C$ corresponds to the value used in \cref{eq:Bvalue}. Usually, the bound-state search takes  1 to 5 iterations, with more iterations needed for energies closer to threshold where $B$ is larger). Once that convergence is reached, the corresponding eigenvector of coefficients $c_{K\gamma i}$ gives a bound state as defined by \cref{eq:fullstate}. 

\subsection{Scattering states} 
For positive energies there is a continuous family of solutions to \cref{eq:blochschrodinger}. At a given energy, the external solutions are given by \cite{Descouvemont2006,Baye2009}
\begin{align}
    &\chi_{K\gamma( K'\gamma')}^{J\pi, ext}(E,\rho)=i^{K'+1}(2\pi/k)^{5/2}\nonumber \\
    &\qquad\left[H^-_{K\gamma}(k \rho)\delta_{\gamma \gamma'}\delta_{K K'}-U^{J\pi}_{K\gamma,\gamma' K'}H^+_{K\gamma}(k\rho)\right]\label{eq:scatteringsol}
\end{align}
where primed indices denote the entrance channel, $k=\sqrt{2m_NE}/\hbar$, $U$ is the scattering matrix and $H^\pm$ are the outgoing and incoming waves 
\begin{equation}
   H_{K\gamma}^\pm(x)=\pm i\sqrt{\frac{\pi x}{2}}\left[J_{K+2}(x)\pm i Y_{K+2}(x)\right],
\end{equation}
 with $J_n$ and $Y_n$  Bessel functions of first and second kind, respectively. 
 
 For scattering states, the boundary value is set to  zero, i.e., $B=0$ in \cref{eqBlochOp}, and we solve \cref{eq:blochschrodinger} by explicitly constructing the $R$-matrix. The $R$-matrix can be obtained from the inverse of the $C$-matrix~\cref{eqCmatrix} as~\footnote{We use a different normalisation of the $R$-matrix from \cite{Descouvemont2010}.}
\begin{equation}
    R^{K\gamma}_{K'\gamma'}(E)=\frac{\hbar^2k}{2m_N}\sum_{ij}\hat{f}_i(\rho_{max})\left\{(C(E)^{K\gamma }_{K'\gamma' })^{-1}\right\}_{ij}\hat{f}_j(\rho_{max})
\end{equation}
where $C(E)$ takes the form in \cref{eqCmatrix} minus the scattering $E$, and $\hat{f}_i$ are the hyperradial mesh functions detailed in Appendix~\ref{AppHyperradialmesh}.
  We can then  obtain the scattering matrix $U$ by solving
\begin{equation}
    Z^+U=Z^-.
\end{equation}
  with \footnote{\cref{eqZexpression} differs from expressions in \cite{Descouvemont2010} by the fact that the $i^{K'+1}$ term is present in both, so $Z^-\neq (Z^+)^*$.}
 \begin{align}Z^{\pm}&=\left(R^{K\gamma}_{K'\gamma'}(E)H^{\pm}_{K\gamma}(k\rho_{max})\right.\\
 &\left.-\delta_{KK'}\delta_{\gamma\gamma'}H^{\pm'}_{K\gamma}(k\rho_{max})\right)i^{K'+1}. \label{eqZexpression}
\end{align}
The internal wavefunction \cref{eq:fullstate}  is  obtained with 
\begin{align}
   & c_{K\gamma i}(E)=\frac{\hbar^2k^{3/2}(2\pi)^{5/2}}{2m_N}\sum_{K'\gamma'}
    i^{K'+1}\nonumber \\
    &\qquad\left(H^-_{K'\gamma'}(k\rho_{max})\delta_{K'K_0}\delta_{\gamma\gamma_0}-U^{K'\gamma'}_{K_0\gamma_0}H^{+'}_{K'\gamma'}(k\rho_{max})\right)\nonumber \\
    &\qquad\sum_{ij}\hat{f}_i(\rho)\left\{(C^{K\gamma }_{K'\gamma' })^{-1}\right\}_{ij}\hat{f}_j(\rho_{max}).
\end{align}
where ${K_0\gamma_0}$ denotes the entrance channel. 

The scattering matrix $U^{J\pi}$ for each spin-parity $J^\pi$ still has a large dimension, and in principle could contain large off-diagonal couplings. At each energy, an eigen-decomposition of the S-matrix allows us to define the eigen-phases $\delta_n(E)$ 
\begin{equation}
    U^{K\gamma}_{K'\gamma'}(E)=\sum_{n}e^{2i\delta_n(E)}v^{(n)}_{K\gamma}(E)v^{(n)}_{K'\gamma'}(E).\label{eq:eigenphase}
\end{equation}
These eigenphase at different energies are computed looking for a continuous evolution with energy of the eigenfunctions $v^{(n)}(E)$. Such approach is useful since the eigenfunctions contain the weights of the individual channels which contribute most strongly to a given structure in the phaseshifts.

\section{Three-body electric multipole operators} 
\label{sec:Be1}
In this work, we are interested in predicting the dipole strength of two-neutron halo systems.
The electric multipole operators are defined by an expansion of the response of the nucleus to electromagnetic fields. In general, they are associated with the strength of transitions between bound states or between a bound  and  continuum states (e.g. in radiative capture or  photodissociation~\cite{Krane1987,BlattWeisskopf1952}). 
The reduced transition strength associated with electric transitions between an initial and final state with spin and parity $J_{i,f}^{\pi_{i,f}}$ is \cite{BohrMottelson}
\begin{equation}
    B(E\lambda)\propto \left|\rme{J_f^{\pi_f}}{\ve{r}^\lambda Y_{\lambda}}{J_i^{\pi_i}}\right|^2
\end{equation}
where $Y_\lambda$ is a spherical harmonic. When the final or initial state is bound, the multipole strengths are strongly influenced by the asymptotics of the three-body bound wavefunction, and hence by its asymptotic normalisation constant, due to the factor of $\ve{r}^\lambda$.

The dipole strength $B(E1)$ is driven by the difference between the centers of mass and charge of the nucleus. It  is enhanced in neutron halos, since the two loosely-bound neutrons can shift the center-of-mass of the nucleus. Using our expansion of the three-body wavefunction~\cref{eq:fullstate} for bound and scattering states (see Sec.~\ref{sec:Rmatrix}, this operator reads for a two neutron halo (taking the particle (3) as the core)
\begin{widetext}
\begin{align}
\frac{dB(E1)}{dE}&=(2\pi)^{-7}\frac{9}{2}\frac{4m_N^3E^2}{\hbar^6}(Z_3e)^2\sum_{J}\hat{J}^2\sum_{K'\gamma'}{\frac{2}{A_3A_{tot}}}\left|\sum_{K\gamma;K_0\gamma_0}\delta_{SS_0}(-1)^{L+S+K} \hat{L_0}\,\hat{L}\,\hat{l}_{x0}\,\hat{l}_{y_0}\right.\sixj{L_0}{J}{S}{1}{J_0}{L}\nonumber\\
&\int d\rho\, \rho \,\chi^{J\pi}_{K\gamma(K'\gamma')}(\rho)\chi^{J_0\pi_0}_{K_0\gamma_0}(\rho)
\left.\int d\alpha\sin^{3}\alpha\cos^{2}\alpha \,\phi_{l_xl_y}^K(\alpha)\phi_{l_{x0}l_{y0}}^{K_0}(\alpha) \begin{Bmatrix}
        0&1&1\\
        l_x&l_y&L\\
        l_{x0}&l_{y0}&L_0
    \end{Bmatrix}
C_{l_{x0}000}^{l_x0}C_{l_{y0}010}^{l_y0}\right|^2\label{eqn:be1}
\end{align}
\end{widetext}
where we use the notation $\hat{J}=\sqrt{2J+1}$, $Z_3$ is the core proton number,  and $A_{tot}=A_3+2$. This is a particular case of a more general expression that is derived in appendix \ref{sec:multipole}. 

Practically, the $\alpha$ integral is done using the same Gauss-Legendre quadrature that was used to compute Hamiltonian matrix elements. The integral over $\rho$ is also evaluated with the Gauss expansion in the region $\rho<\rho_{max}$,  the ANC of each channel is calculated and a Gauss-Lagrange integration is performed with the exact expression \cref{eq:boundext} from $\rho_{max}$ to $\infty$. Except at very low energies, the external integral generally has a negligible effect, since the channel radius must already be very large in order to ensure convergence (see Appendix~\ref{AppConvergence}).

\section{Results}
Having introduced the formalism and method used to describe two-neutron halo systems, we focus on the case of $^{22}$C, expected to be a two-neutron halo \cite{Tanaka2010,Kobayashi2012}. Since $^{21}$C is unbound to neutron emission \cite{Mosby2013},  $^{22}$C is a Borromean halo nucleus, i.e. it does not contain any  bound  two-body subsystem. Previous theoretical studies of this nucleus \cite{Horiuchi2006,Pinilla2016}, relying on supersymmetry to remove the Pauli-forbidden states, suggest a dominant $(\nu 1s_{1/2})^2$ ground state configuration. In this work, we will apply our formalism to investigate how the choice of method to remove Pauli-forbidden states influences the bound properties of $^{22}$C, namely its binding energy, r.m.s. hyperradius, partial-wave decomposition, Dalitz plots, and dipole strength.

\subsection{Choice of $^{22}$C and interactions} 
\label{sec:potentials}
\begin{table}
    \centering
    \begin{tabular}{cccc|c}
         $n$&$l_a$&$[I_1,I_3]s$&$[l_a,s],j$& E (MeV) \\ \hline 
        0& 0&1/2&1/2&-17.04 \\
         0&1&1/2&3/2&-9.98 \\
       0& 1&1/2&1/2& -5.05\\
      0& 2&1/2&5/2& -1.63
    \end{tabular}
    \caption{Quantum numbers and energies of the forbidden states in the core-neutron system. Note that the sum of their multiplicities adds up to 14, the number of neutrons in the $^{20}$C core.}
    \label{tab:forbiddenstates}
\end{table}

In our framework,  we need as input a $^{20}$C-$n$ potential as well as some $n$-$n$ interaction. In this section, we detail how these interactions are constrained. 
Similarly to Refs.~\cite{Tanaka2010,Pinilla2016}, we use the Minnesota potential \cite{Bogner2011} for the $n$-$n$ interaction in the $s$-wave, and no interaction in all other $n$-$n$ partial waves. 

 The $^{20}$C-$n$ interaction is simulated  with a Woods-Saxon potential defined for each partial wave by a radius $R_l$, depth $V^0_l$ and diffuseness $a_l$ as
\begin{align}
    V_l^{WS}(r)&=-V^0_l f_{WS}(r),\\
   \text{with}\quad f_{WS}(r)&=\frac{1}{1+\exp((r-R_l)/a_l)}.
\end{align} We also include a spin-orbit potential of the form 
\begin{equation}
    V_l^{so}(r)=\frac{V_l^{so}}{m_\pi^2} \,\ve{l}\cdot\ve{s}\,\frac{1}{r}\, \frac{df_{WS}(r)}{dr}
\end{equation} where $m_\pi$ is the pion mass. 
 The values of the two-body parameters (see Table \ref{tab:parameters}) are chosen to reproduce two-body scattering quantities consistent with the (limited) known properties of $^{21}$C. In particular, this choice of parameters gives an s-wave scattering length $a_s=-2.50$~fm, consistent with that inferred from experimental data~\cite{Mosby2013}. It also places the bound $d_{5/2}$ state at $-1.62$~MeV, corresponding to the $l=2$ peak seen in neutron knockout spectra from $^{21}$C \cite{Leblond2015}. The quantification of uncertainties associated with the adjustment of these parameters on two-body data will be the subject of a future work \cite{UQpaper}.
With the core-neutron interaction considered here, there are four forbidden states which must all be removed; their energies and quantum numbers are given in Table~\ref{tab:forbiddenstates}.

The next section will study how the choice of the method to remove these forbidden states impact the properties of of $^{22}$C.
When only considering the two-body forces, this system has no bound state, regardless of the method to remove the Pauli-forbidden states. To produce a bound state but remove the effects of the binding energy (that is often enforced in three-body models~\cite{Casal2020,Pinilla2025,Esbensen2007}),  we fix it to {-0.5}~MeV by introducing a three-body force (see Section~\ref{Sec:3bforce}) with a depth of $-14.36$~MeV for the projection case and $-14.93$~MeV for the  supersymmetry one. All results in this manuscript use this three-body force in $J^\pi=0^+$ channels, while we do not consider any  in other channels.

\subsection{Comparison of supersymmetry and projection approaches}\label{sec:resultsSusyProjection}

\begin{figure}
    \centering
    \includegraphics[width=\linewidth]{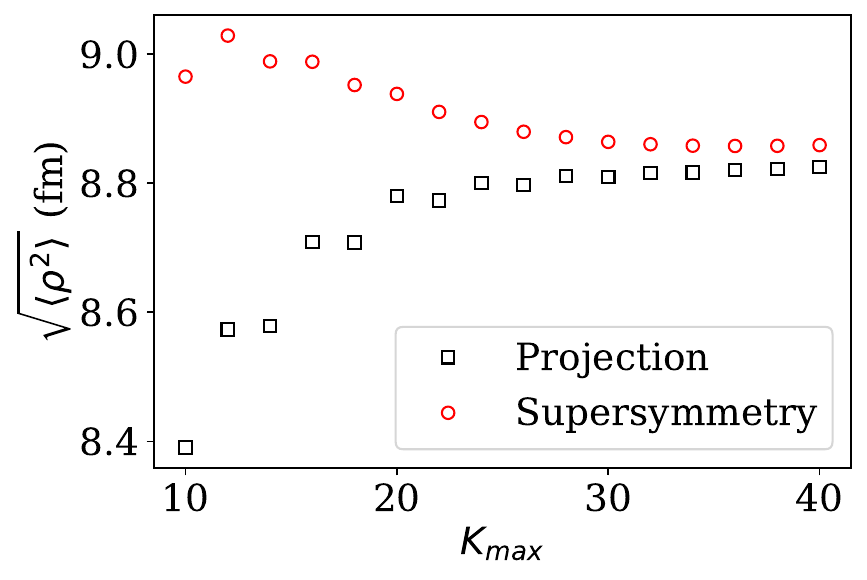}
    \caption{Root-mean-square  hyperradius of bound states calculated with different $K_{max}$ cutoffs.}
    \label{fig:rmsbound}
\end{figure}

As shown in Appendix~\ref{AppConvergence}, the cutoffs $K_{max}=40$, $\rho_{max}=60$ and $N=60$ lead to converged results. Except when specified otherwise, all calculations shown in this section use this model space. Interestingly the different methods of removing forbidden states have similar smooth convergence patterns for the three-body binding energy, i.e., they decrease almost exponentially with $K_{max}$ (see Appendix~\ref{AppConvergence}).

We first study in Fig. \ref{fig:rmsbound} the convergence  of the r.m.s. hyperradius of the $0^+$ ground state of $^{22}$C.  Interestingly,  both  methods reach convergence within 0.02~fm at $K_{max}=30$ although their convergence patterns differ for low $K_{max}$. 
At  $K_{max}=30$, both bound states have approximately equal binding energy and  r.m.s. hyperradii (8.82~fm with projection vs. 8.86~fm with supersymmetry). These small differences suggest that perhaps both methods are equivalent. To further compare the two, we next consider the spectroscopic decomposition of the three-body wavefunction into the different partial waves.

\begin{figure}
    \centering
    \includegraphics[width=\linewidth]{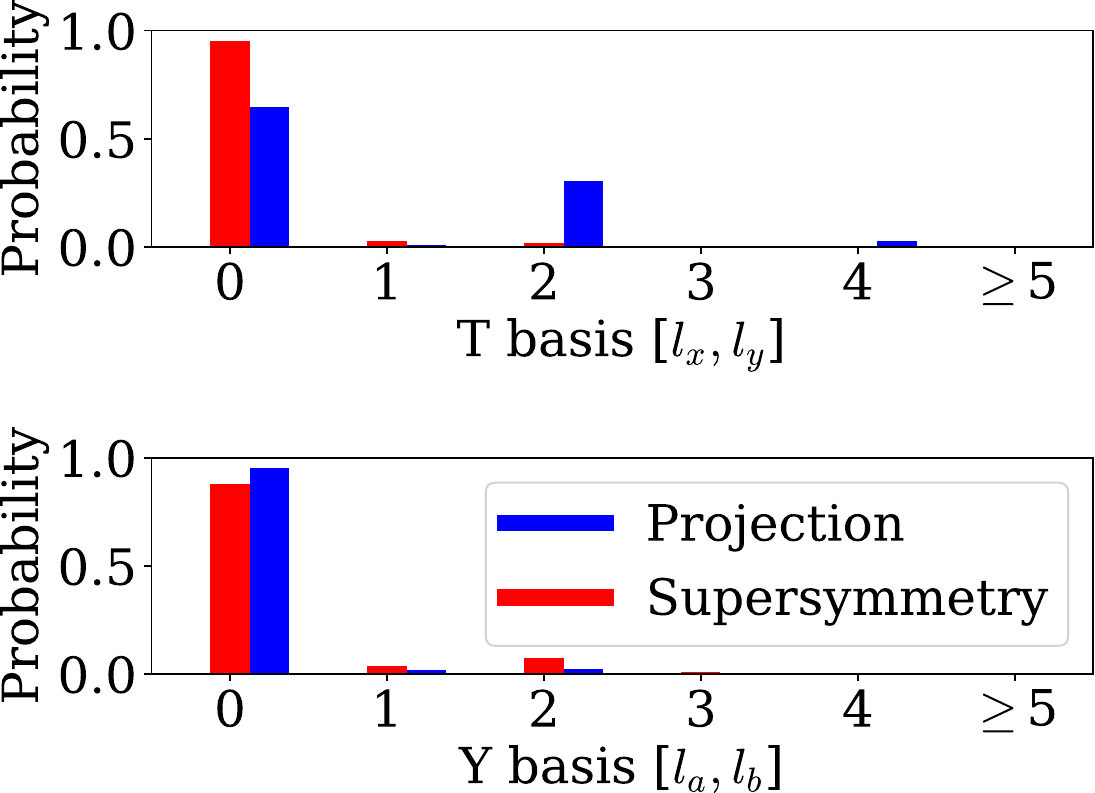}
    \caption{Amplitudes of the lowest angular momentum partial waves in the $0^+$ ground state at $-0.5$~MeV calculated with projection and supersymmetry. $l_x,l_y$ angular momenta are in the T-basis and $l_a,l_b$ angular momenta are in the Y-basis.}
    \label{fig:occupations}
\end{figure}

Figure~\ref{fig:occupations} shows the dominant partial waves in the T-basis (top) and Y-basis (bottom) of the $0^+$ ground states obtained with both methods. Because the core-neutron interaction has a longer range,  it is more natural to  first examine the Y-basis  to  understand why the supersymmetry method leads to slightly larger r.m.s. hyperradius for the ground state. In the Y-basis, the amplitudes are  similar, the only difference being that the supersymmetry calculations lead to the population of higher-$l$ partial waves. These are subject to larger centrifugal barriers, and will have more spatially extended components, providing  an explanation of why the radii predicted by supersymmetry are slightly larger than in the projection case. This population of higher-partial wave is likely due to the strong repulsive short-range potential introduced by the supersymmetry method.
In the T-basis, these differences translate into to $\sim$30\% less d-wave component and $\sim$30\% more s-wave component for the supersymmetry compared to the projection method.

\begin{figure}
    \centering
    \includegraphics[width=\columnwidth]{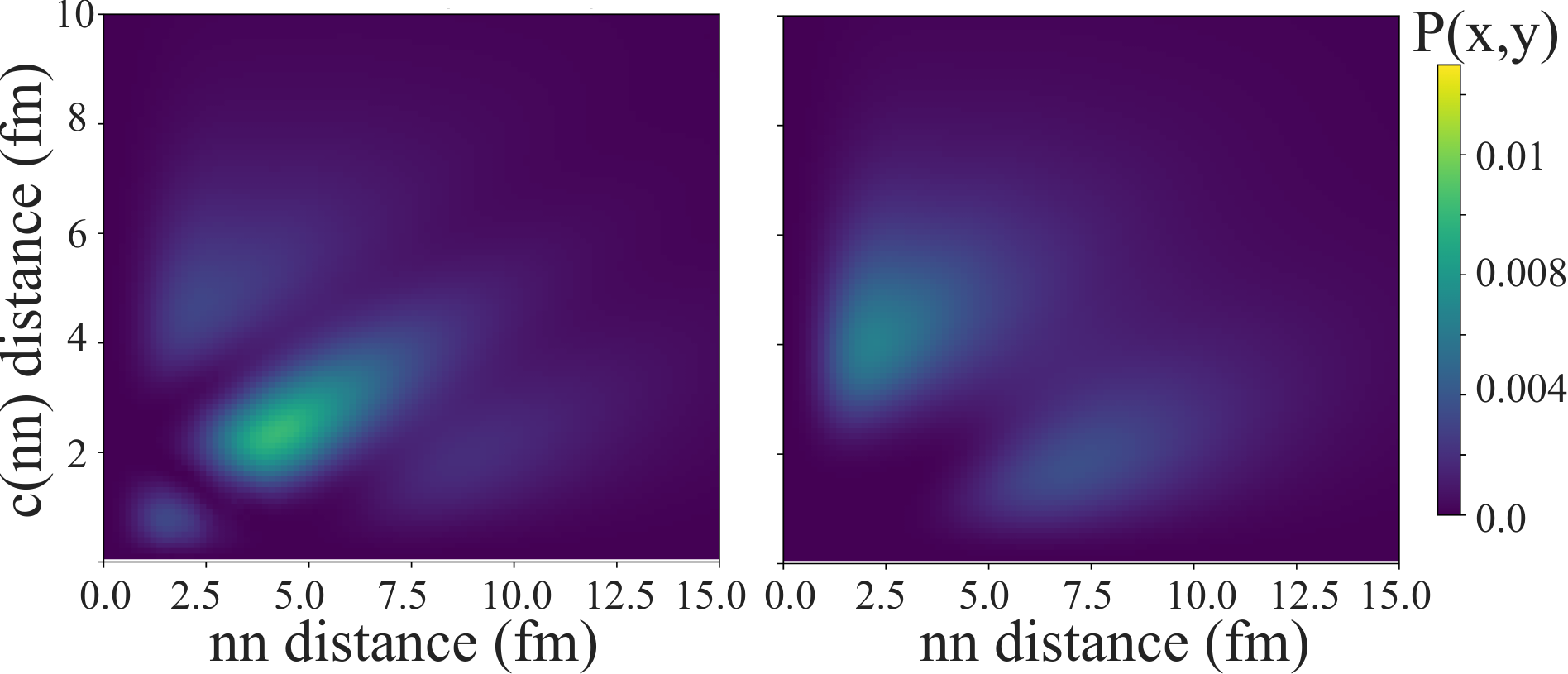}
    \caption{Dalitz plots of the ground state solutions with $K_{max}=30$ for projection (left) and supersymmetry (right). Note that $r_{nn}=x/\sqrt{\mu_{nn}}$}
\label{fig:Dalitz}
\end{figure}

We now study the probability to find the system with some distances (losing information about relative orientations), defined as
\begin{equation}
    P(x,y)=\int d\Omega_x\int d\Omega_y \, x^2 y^2 |\Psi(\mathbf{x,y})|^2.
\end{equation} 
Figure~\ref{fig:Dalitz} are Dalitz plots showing that these probability distributions for both methods are vastly different. In the projection case (left), the most likely configuration has the neutrons separated by $\sim$5~fm at about 2.5~fm from the core. It  also has some possibility of more pronounced ``cigar" (small $c$-$(nn)$ and large $n$-$n$ distances) or ``di-neutron" configurations (large $c$-$(nn)$ and small $nn$ distances), as well as a configuration with both neutrons close and near the core (small $c$-$(nn)$ and  $n$-$n$ distances). The supersymmetry solution, however, is dominated by the di-neutron configuration, with some mixture of the cigar. The differences in the Dalitz plots obtained woth  both methods are similar to that seen in a comparison of the same two methods applied to $^{12}$C in Ref.~\cite{Descouvemont2003}.

\begin{figure}
    \centering
    \includegraphics[width=0.9\columnwidth]{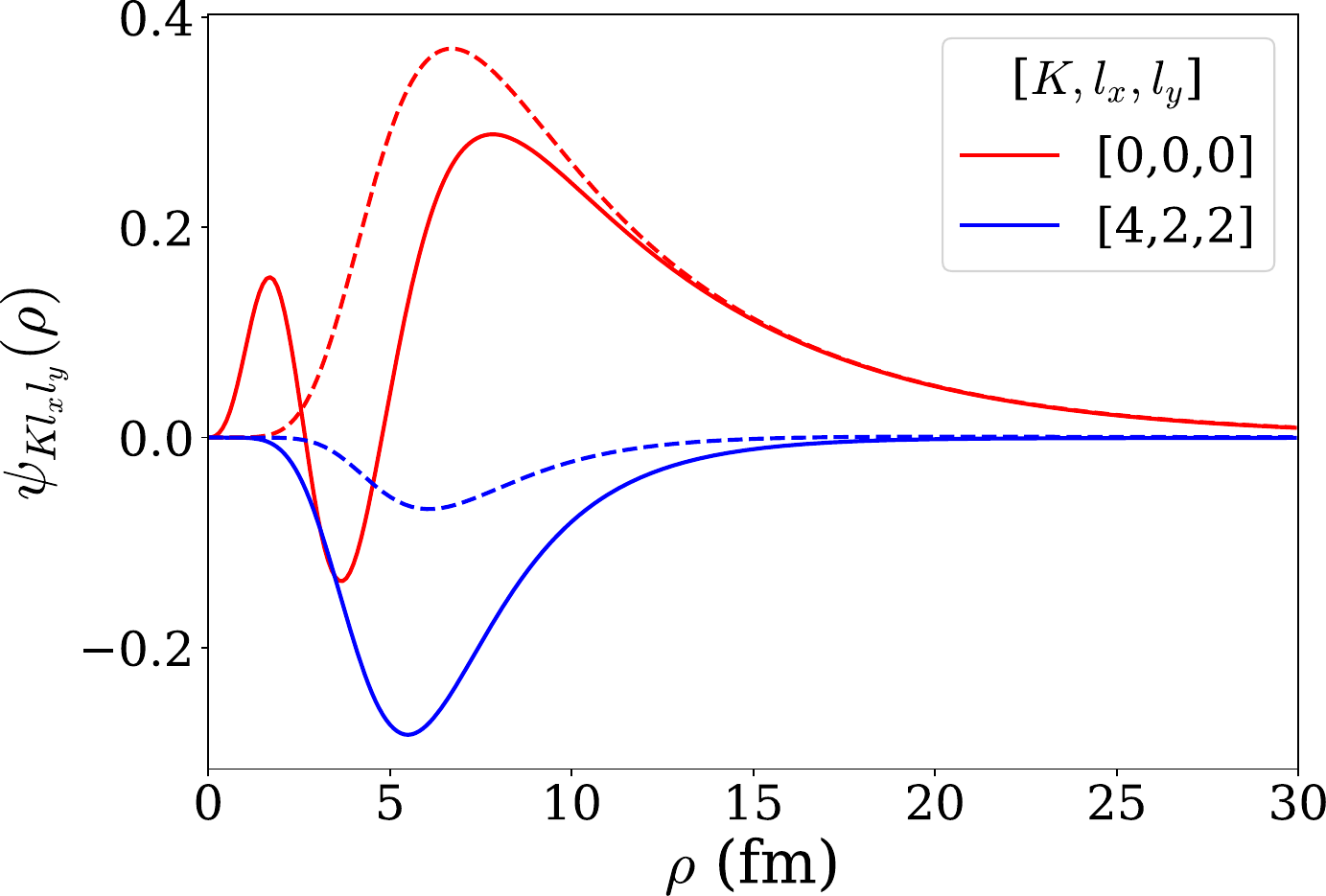}
    \caption{Wavefunctions of the two largest partial waves in each case, projection plotted with solid lines and supersymmetry with dashed. The total amplitude of the chosen partial waves for projection (supersymmetry) are: 0.54(0.83), 0.30(0.02)}
    \label{fig:wavefunction}
\end{figure}

From the analysis of bound-state properties, it is clear that projection and supersymmetry methods do not predict the same physics. To understand which one is more accurate,  we compare in Fig.~\ref{fig:wavefunction} the radial wavefunctions of the partial waves with the largest amplitudes in the ground state predicted using the projection (solid lines) and supersymmetry  (dashed lines) methods. As already shown in Fig.~\ref{fig:occupations}, both methods lead to different normalization of each component. Interestingly, they also differ at short hyperradius, i.e. the wavefunctions obtained with the projection method exhibit  extra nodes compared to the supersymmetry case. These nodes can be explained by a proper enforcement of the Pauli principle which means that the three-body wavefunctions are correctly anti-symmetrised  with the neutrons in the $^{20}$C core. Similar results were seen in Ref. \cite{Thompson2000}, where introduction of Pauli projection resulted in additional nodes in the $K=0$ channel, but did not affect the $K=2$, which was the dominant hypermomentum in $^{6}$He. In contrast to that study which involves a $p$-wave dominated halo, $^{22}$C has its valence neutrons in the $sd$ region, so the effect of Pauli projection is significant as it redistributes ground state probability from the $K=0$ configuration which only includes $l_x,l_y=0$, to, in this case, the $K=4$ configuration which comports $l_x,l_y\leq 2$.

Finally, we now discuss the impact of the treatment of the Pauli-forbidden states on the dipole strength of $^{22}$C in Fig.~\ref{fig:convergenceBE1}. Similarly to the phase shifts (see Appendix~\ref{AppConvergence}), the dipole strength computed with supersymmetry (dashed lines) converges at lower $K_{max}$ than the one obtained with the projection method. Interestingly, in both cases, we can see that the peaks are shifting to lower energy for larger $K_{max}$. Since there is no three-body force applied to the $J^\pi=1^-$ scattering waves, it is expected that features of the continuum to move towards lower energy when  increasing the model space. Although the shapes are similar, the magnitudes differ almost by a factor 2 at the peak. This difference is largely to the difference in the occupation of the various partial waves and hence the different ANCs calculated with each method. It appears that the transition strength is dominated by capture to the $[l_x,l_y]=[0,0]$ partial wave in the T-basis. Therefore the strength of the $B(E1)$ with projection is suppressed by a factor of ~0.58, very similar to the ratio in Table~\ref{tab:occupations}. 
\begin{figure}
    \centering
    \includegraphics[width=\linewidth]{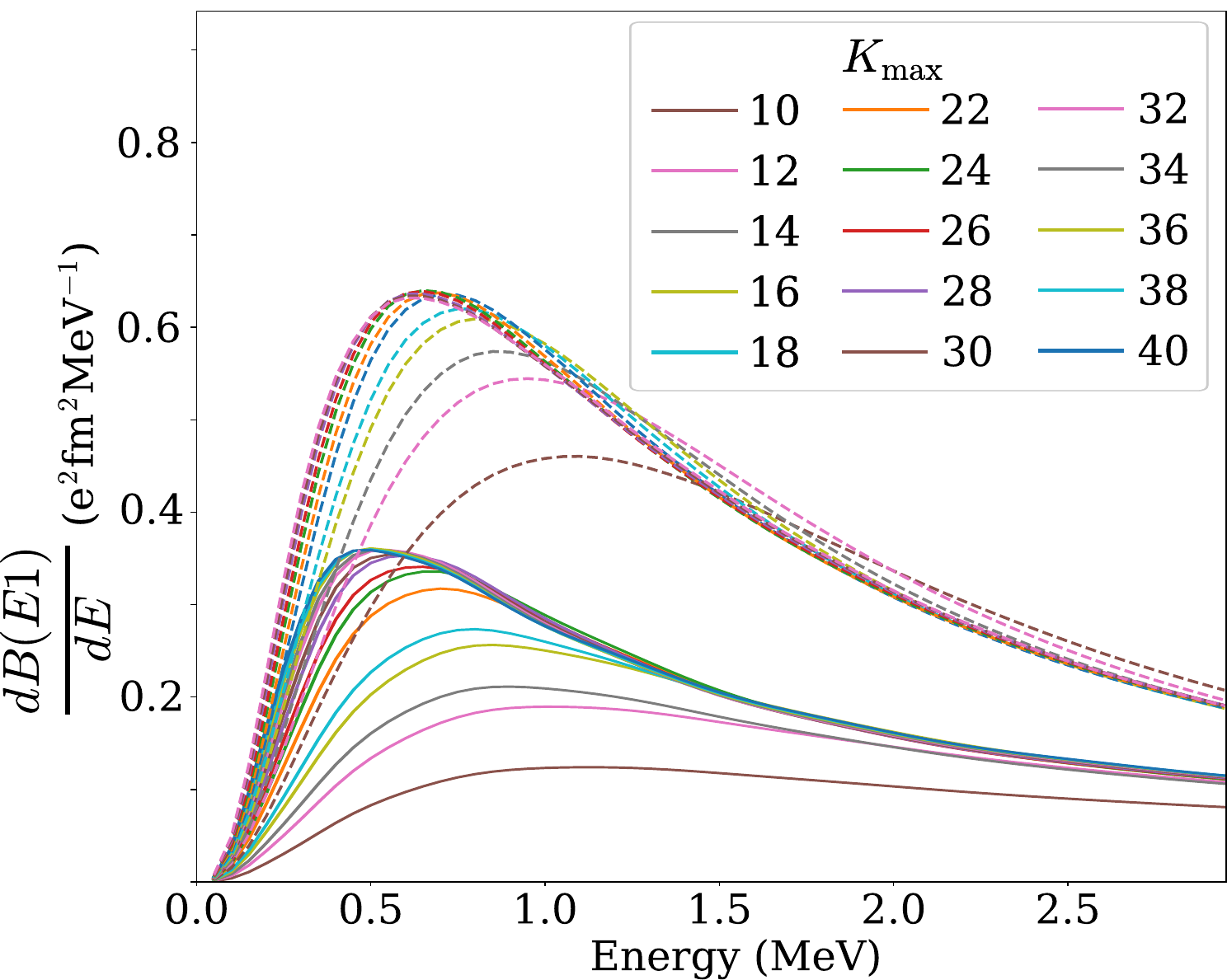}
    \caption{$B(E1)$ strength functions calculated with different cutoffs in $K_{max}$. All calculations are with $\rho_{max}=60$~fm,$N=60$ and a three-body force adjusted to reproduce a $0.5$~MeV binding energy. Solid lines are calculated with projection and dashed lines with supersymmetry. Note that supersymmetric calculations reach convergence at $K_{max}=32$ so there are fewer curves.}
    \label{fig:convergenceBE1}
\end{figure}

 Our detailed analysis shows that the ground- and scattering-state properties of $^{22}$C as well as its dipole strength are impacted by the choice of the method  to enforce Pauli-exclusion principle in this three-body model. It emphasizes that the supersymmetry method to remove Pauli-forbidden states, although simpler computationally and algorithmically, is not accurate and that the projection method should be favored.

\section {Speeding up calculations} \label{Sec:Speedup}
With the goal of quantifying the uncertainties associated with the  potential parameters in three-body observables, we discuss in this section various avenues to decrease the computational cost of one individual calculation. The computation time can generally be expressed in terms of $\mathcal{O}(d^n)$ notation, where $d$ is the dimension of the Hamiltonian for the largest $J^\pi$ channel considered, where $d=N\times n_\gamma$, and $n_\gamma$ is the number of partial waves in that channel. Some tasks, especially diagonalisation and inversion, are costly and scale poorly with $d$ ($\sim\mathcal{O}(d^3)$), so the trade-off between speed and convergence is important. The use of modern computing architecture and especially parallelisation is paramount to speeding up calculations. In our implementation, we have distributed the computation of the Hamiltonian across independent processes using \texttt{MPI}, and different $J^\pi$ channels are computed entirely independently. Operations which require the entire Hamiltonian are the main bottlenecks of the code. However, these tasks are  accelerated by the use of multithreading with \texttt{openMP}.

\begin{figure}
    \centering
    \includegraphics[width=\linewidth]{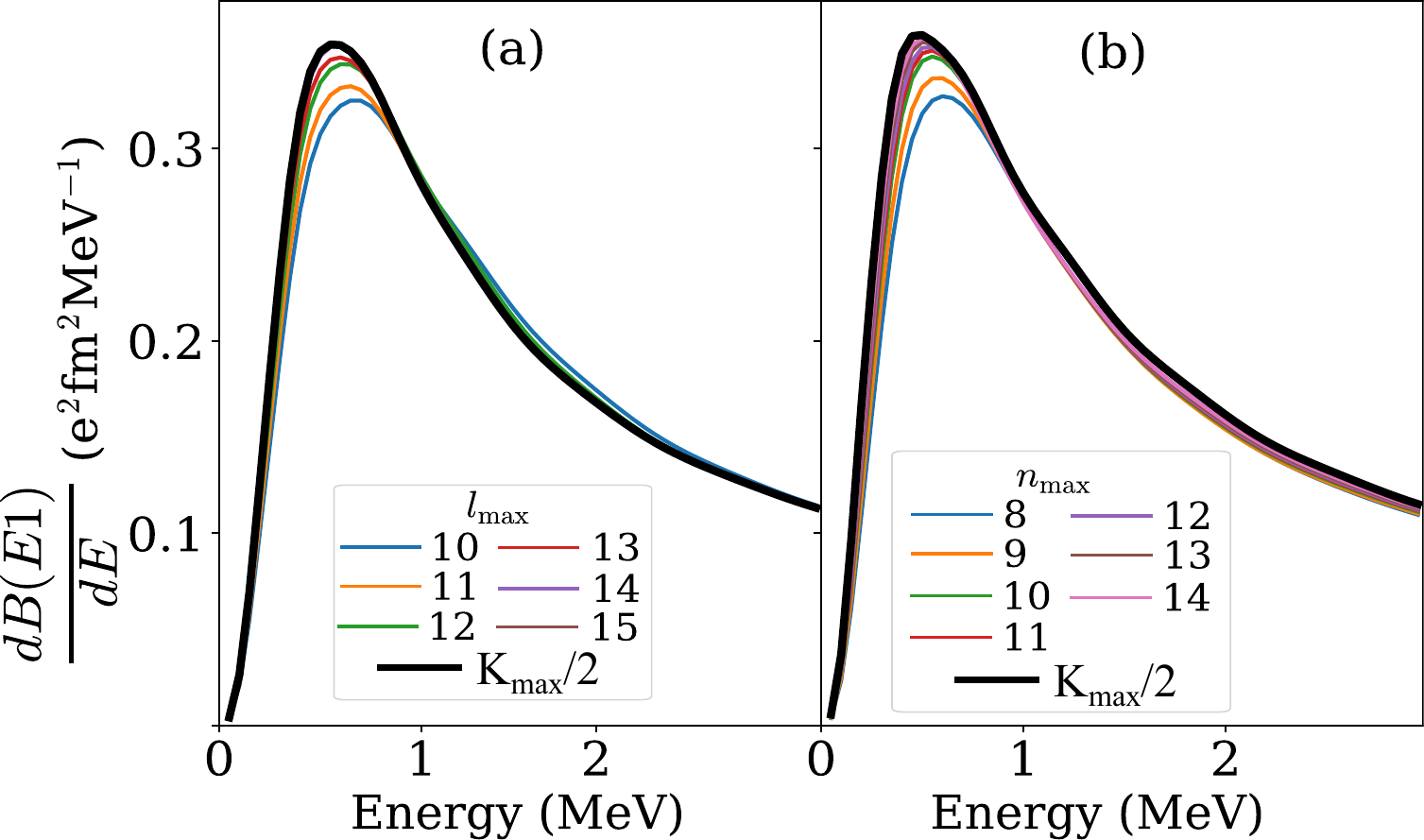}
    \caption{$B(E1)$ strength functions calculated with different cutoffs in (a) $l_{max}$ and (b) $n_{max}$. All calculations are with $\rho_{max}=60$~fm,$N=60$ and projection. $l_{max}$ truncation is done with $K_{max}=30$ and $n_{max}$ truncation is with $K_{max}=40$.}
    \label{fig:cutoffs}
\end{figure}

Besides an efficient implementation, we also investigate different truncations of the number of partial waves, that enable descreases in $d$ and hence the computational cost. The number of partial waves with a given value of $K$ increases with $K$ so that $d\sim K_{max}^2$, which means adding cutoffs on unimportant partial waves for high $K$ can be helpful. We use two such cutoffs: $n_{max}$ limits the maximum value of $n=\frac{K-l_x-l_y}{2}$ (and therefore has no effect when $n_{max}\geq K_{max}/2$) and $l_{max}$  limits the maximum values of both $l_x$ and $l_y$ individually (and therefore has no effect when $l_{max}\geq \lceil K_{max}/2\rceil$). As $n_{max}$ acts to remove small $l$ partial waves with large $K$ while $l_{max}$ removes large $l$ partial waves, it is not desirable to use cutoffs in both $l_{max}$ and $n_{max}$ simultaneously since this amounts to an effective $K_{max}$. 

\begin{figure}
    \centering

    \vspace{0.5cm}
    \includegraphics[width=0.9\linewidth]{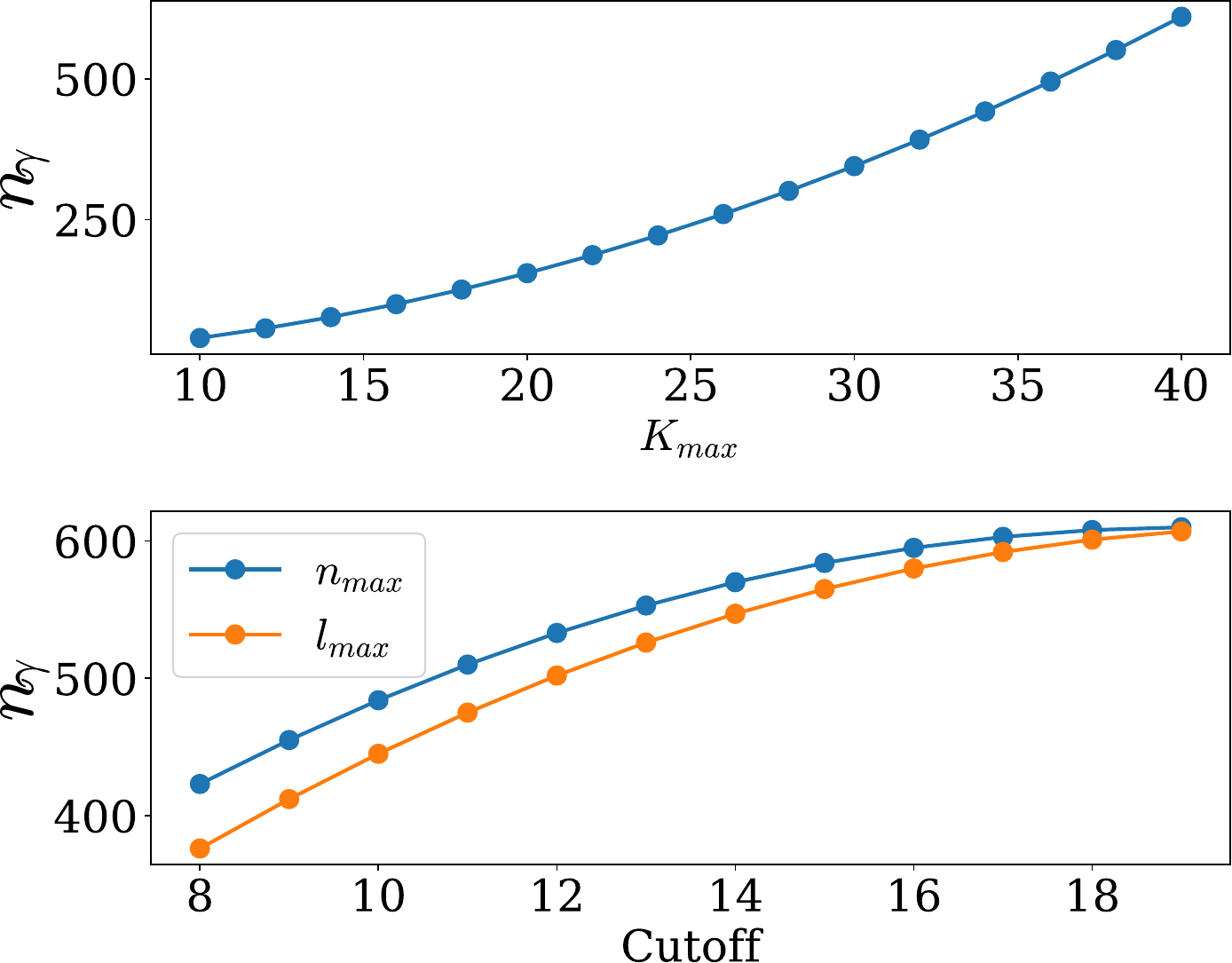}
    \caption{(above) Number of partial waves in the $J^\pi=1^-$ channel for different values of $K_{max}$. (below) Number of partial waves in the same channel for different cutoffs in $n_{max}$ and $l_{max}$, with $K_{max}=40$.}
    \label{fig:numpartials}
\end{figure}

Figure~\ref{fig:cutoffs} shows the effect of cutting off high $n$ or $l$ channels for $B(E1)$ strength obtained with the projection method. It demonstrates that $l_{max}$ is a poor cutoff, since it has to be extended to nearly its maximal value ($K_{max}/2$) to obtain convergence. On the other hand, $n_{max}$ can be set lower and still reproduce the full calculation, i.e., the one with no cutoff in $l_{max}$ or $n_{max}$.  Figure~\ref{fig:numpartials} shows the scaling of the number of the partial waves as a function of the different cutoffs. Only the number of partial waves in the $1^-$ (scattering) channel are shown here, since the bound $0^+$ computation is done in parallel and never takes longer than the scattering state~\footnote{For $K_{max}=40$, the bound state has 231 partial waves compared with 610 for the scattering state.} Recalling that the total number of CPU-hours is approximately $\mathcal{O}((N\times n_\gamma)^3)$, even the small gain in dimension from $n_{max}=14$ compared with $n_{max}=20$ corresponds to a $\sim$20\% CPU time reduction.

\section{Conclusions}\label{Sec:Conclusions}

 Loosely bound nuclei, or those located far from stability, exhibit a transition from many-body degrees-of-freedom to effective few-body ones, manifesting itself through the emergence of cluster structures. This fascinating transition motivates the development of   theoretical models to treat accurately both the shell effects driving the single-particle structure, as well as the few-body dynamics between these emerging clusters.  In this work, we adopt a few-body model, as in Refs~\cite{Pinilla2016,Descouvemont2006,Pinilla2011}, in which coupling to continuum states is included and the shell structure is enforced through the use of effective interactions and operators to enforce the Pauli-exclusion principle.  We first demonstrate the  capabilities of our new code to predict the properties of $^{22}$C and we compare  two  methods to remove Pauli-forbidden states, i.e. the supersymmetric and projection methods. We show that  the choice of method strongly impacts  the partial-wave content of the ground state wavefunction as well as its dipole response. By studying the differences in their  wavefunctions, we argue that the projection method is more accurate and should be favored. This suggests that it would be interesting to revisit predictions of two-neutron halo nuclei that relied on supersymmetric transformation, to verify if the predictions remain unchanged when the projection method is used to remove Pauli-forbidden states.

This work also presents numerous technical developments: it details how modern parallelised architecture can be leveraged to speed up the computations, it also introduces new algorithms to apply the projection method and to adjust three-body force, as well as providing general theoretical derivations of electric multipole strength in three-body hyperspherical harmonics formalism. Using our new implementation, we also determine how the computational cost can be further reduced by exploring  truncations in number of three-body channels. These important developments pave the way for robust uncertainty quantification in few-body models, treating consistently bound and scattering properties of two-neutron halo systems, without using a discretization of the three-body continuum. Moreover,  this code provides a useful basis  to predict more complex systems, e.g. three-body charged systems, and/or a more complete set of observables. Both extensions are parts of our future plans.

\begin{acknowledgments} 
We are grateful to Filomena Nunes with whom we had frequent discussions about the implementation of this formalism, and to the few-body reaction group at MSU. We also thank Pierre Descouvemont to have provided values to benchmark our code, as well as Guillaume Hupin and Mark Caprio for interesting discussions about the development of the code and its optimization. Calculations were performed using the High-Performance Computing Center at MSU's Institute for Cyber-Enabled Research. This project has received financial support from the CNRS through the AIQI-IN2P3 project.
\end{acknowledgments}

\bibliography{sources}
\appendix 

\section{Raynal-Revai transformation of Hamiltonian} \label{sec:RRtransform}

Because of our choice of spin coupling scheme \cref{eq:coupling}, the transformation between Jacobi sets has three steps: decoupling spins $I$ from orbital angular momenta $l$, applying the Raynal-Revai transformation to the orbital angular momenta, and recoupling the spins in the other set. The spin decoupling and recoupling is done following Ref.~\cite{Varshalovich1988} and results in a number of Wigner 6-$j$ and 9-$j$ symbols. Ultimately, it reads 
\begin{align}
   \mathcal{T}_{\gamma'_Y}^{\gamma_T}=&\bra{\gamma'_Y}\ket{\gamma_T}\\
   =&\sum_{j_2}(-1)^{l_y+j^{(T)}+l'_b+j'+3S'+3I_T+2I_3+2J'+2I_2} \nonumber\\
&\quad(\hat {j}_2)^2\,\hat {j} \,\hat {L}\,\hat {S}\,\hat {I_T}\,\hat {j^{(T)}}\,\hat {L^{(T)}}\quad {}_{Kj_2}RR_{l_xl_y}^{l_al_b}\nonumber \\
&\quad\sixj{I_T}{l_y}{l_x}{L^{(T)}}{j^{(T)}}{j_2}\sixj{S}{l_b}{l_a}{L}{j}{j_2}\begin{Bmatrix}
       j_2&S&L\\
       I_T&I_1&I_2\\
       L^{(T)}&I_3&J
   \end{Bmatrix}
\end{align}
where we used the notation $\hat j=\sqrt{2j+1}$, $_{Kj_2}RR_{l_xl_y}^{l_al_b}=\bra{l_a,l_b}\ket{l_x,l_y}_{Kj_2}$ is the Raynal-Revai coefficient \cite{Raynal1970}, which we compute using \cite{Youping1987}.

\section{Variationally exact three-body projection of two-body forbidden states} \label{sec:projection}

The operator which projects the three-body space into the subspace of a particular two-body forbidden state (we label this by its radial wavefunction $u(x)$) is 
\begin{equation}\hat{P}=\sum_{n}\ket{u(x);g_n(y)}\bra{u(x);g_n(y)}\end{equation}
where $\{g_n\}$ forms a basis for $y$ i.e.
\begin{equation}\sum_n\ket{g_n(y)}\bra{g_n(y)}=\hat{\mathbbm{1}}_y.\end{equation}For simplicity we omit spin degrees of freedom, though in practice these must also be summed over with appropriate couplings.
We are interested in matrix elements of the projection operator between basis elements $\Phi_{K\gamma i}=\hat{f}_i(\rho)\phi^K_{l_xl_y}(\alpha)$
where $\hat{f}_i$ are the Lagrange functions. 
\begin{equation}
    P^{K\gamma i }_{K'\gamma'i'}=\sum_{n}\bra{\Phi_{K\gamma i}}\ket{u(x);g_n(y)}\bra{u(x);g_n(y)}\ket{\Phi_{K'\gamma' i'}}\label{eqB3}
\end{equation}
The three-body state $\ket{u(x);g_n(y)}$ can be expressed in terms of its $K,\gamma$ components and their hyperradial wavefunctions
\begin{align}\ket{u(x);g_n(y)}&=\sum_{K\gamma}\varphi_{K\gamma n}(\rho)\ket{\rho;K,\gamma}
\end{align}
with
\begin{align}
    \varphi_{K\gamma n}(\rho)&=\frac{1}{\rho}\int d\alpha \sin^2\alpha\cos\alpha u(\rho\cos\alpha)g_n(\rho\sin\alpha)\phi^K_{l_xl_y}(\alpha).\label{eq:varphi}
\end{align}

Using these hyperradial wavefunctions, we can compute the bra and the ket  appearing in \cref{eqB3} as
\begin{align}
    &\braket{\Phi_{K\gamma i}}{u(x);g_n(y)}=\sum_{K'\gamma'}\left(\int_0^\infty d\rho\hat{f}_i(\rho)\varphi_{K'\gamma' n }(\rho)\right.\\
    &\left.\int_0^{\pi/2}d\alpha \sin^2\alpha\cos^2\alpha \,\phi^K_{l_xl_y}(\alpha)\phi^{K'}_{l_x'l_y'}(\alpha)\right)
\end{align}
 Recognising that the last part of those integrals is a defined orthogonal relation, we simplify that to
\begin{align}
    \bra{\Phi^{K\gamma i}}\ket{u(x);g_n(y)}&\equiv \varphi_{K\gamma ni}\\
    &=\int_0^\infty d\rho\, \hat{f}_i(\rho)\varphi_{K\gamma n }(\rho)
\end{align}
Finally, we obtain the matrix elements
\begin{equation}
    P^{K\gamma i}_{ K'\gamma'i'}=\sum_{n s}\varphi_{K\gamma n i} \,\varphi_{K'\gamma' n i'}
\end{equation}
The accuracy of this method relies on precise computations of 
\begin{align}
    \varphi_{K\gamma n i}=&\int_0^\infty d\rho\hat{f}_i(\rho)\int d\alpha \sin^2\alpha\cos\alpha \\
    &\frac{u(\rho\cos\alpha)}{\rho}g_n(\rho\sin\alpha)\phi^K_{l_xl_y}(\alpha). \label{eq:rhoalphaint}
\end{align}
To achieve this, we define $g_n$ as the basis of Lagrange-Laguerre functions which are orthogonal over $(0,\infty)$, and we use the full expression \cite{Baye2015} for $\hat{f}_i$. We also use a two-body Lagrange mesh method to compute $u(x)$, the forbidden states. The integral over $\rho$ in \cref{eq:rhoalphaint} is done numerically on a Lagrange-Laguerre mesh with a larger number of points than the number of functions used in the expansion $g_n$. The alternative approach is to apply the Gauss expansion which would simplify the integral \cref{eq:rhoalphaint}  to 
\begin{equation}
    \int d\rho\,  \hat{f}_i(\rho)\varphi_{K\gamma n}(\rho)\approx \varphi_{K\gamma n}(\rho_i) w_i^{1/2}
\end{equation}
with $w_i$ the Lagrange-Legendre weights. Performing the numerical integral explicitly multiplies the computational cost by the number of points used in the Laguerre mesh for greater accuracy. The integral over $\alpha$ also uses a Lagrange-Legendre mesh. In practice, convergence of the projection operator is achieved with much fewer than the 100 points  in the $\rho,\alpha$ integrals, and 40 basis functions in the expansion $g_n$. These numbers are deliberately conservative to avoid needing to modify them in future cases, although a test of convergence with the number of basis functions is a sensible check. 
\section{Hyperradial mesh in the internal region}\label{AppHyperradialmesh}
We employ the Lagrange-mesh method \cite{Baye2015}, with Lagrange-Legendre basis functions regularised by a power of $3/2$ to resolve the singularity at the origin. This basis is applicable over a finite range $[0,\rho_{max}]$ where we will solve numerically, for $\rho>\rho_{max}$ we treat the off-diagonal couplings as zero and will use the $R$-matrix approach  to match the two solutions at the boundary. With the Lagrange expansion, the wavefunction is expressed as 
\begin{align}
    \Psi^{JM\pi}&=\rho^{-5/2}\sum_{K=0}^\infty \sum_{\gamma} \sum_i^N c_{K\gamma i}\Phi_{K\gamma i}(\rho,\Omega_5)\label{eq:fullstate}\\
    \Phi_{K\gamma i}&=\hat{f}^{(N)}_i(\rho)\mathcal{Y}_{K\gamma}^{JM}(\Omega_5)
\end{align}
where $\hat{f}^{(N)}_i(\rho)$ is the $i$-th Lagrange-Legendre function of order $N$. In general, each of the cutoffs $K_{max}$, $\rho_{max}$ and $N$ should be taken to a large enough value to see convergence in all observables. This convergence is verified and discussed in detail in  Appendix~\ref{AppConvergence}. 

The Gauss expansion ensures that local potentials remain local in the Lagrange-Legendre basis, i.e.
\begin{equation}
    V_{ij}=V(\rho_i)\delta_{ij}
\end{equation}
where $\rho_i$ is the $i$-th point in the Lagrange mesh. The expansion is also convenient since it gives  analytic forms for the kinetic and Bloch operators (see Ref. \cite{Baye2015}). 

\section{Adjustment of three-body force to reproduce three-body binding energy}\label{Sec:3bforce}

For simplicity and due to the absence of a well-established nucleus-neutron-neutron three-body potential model, we use a simple and local form of the three-body force $V^{(3b)}$ and we adjust its strength to reproduce the experimental three-binding threshold, i.e., the two-neutron separation energy in two-neutron halo nuclei. This force reads
\begin{equation}
    V^{(3b)}_{K\gamma i,K'\gamma'i'}(\xi)= \xi e^{-\rho_i^2/R^2}\delta_{KK'}\delta_{\gamma\gamma'}\delta_{ii'}
\end{equation}
$R$ is a range parameter which we fix in this work at 5~fm. In general its effect should be small \cite{Casal2020}. 
We do not consider a three-body force in all partial waves where there is no bound state.  

In order to automate the search for the $\xi$ value that reproduces the binding energy,  we use an iterative procedure. The idea is to update the value of the strength of the three-body force $\xi_j$ at each iteration, and compute the three-body binding energy $E_j$ corresponding to this Hamiltonian $\mathcal{H}_j$.
 The value of the strength of the three-body force  at the $j^{th}$ iteration, $\xi_j$, is computed using perturbation theory and solving
\begin{equation}
    E_{j}=E_0+\frac{dE_{j-1}}{d\xi_{j-1}}(\xi_j-\xi_{j-1})+\frac{1}{2}\frac{d^2E_{j-1}}{d\xi_{j-1}^2}(\xi_j-\xi_{j-1})^2,\label{eq30}
\end{equation}
 where $E_0$ is the target binding energy. The derivatives appearing in \cref{eq30} are obtained as
\begin{align}
    \frac{dE_j}{d\xi_j}&=\bra{\chi^j _0}(1-\mathcal{P})V^{(3b)}(1)(1-\mathcal{P})\ket{\chi^j_0}\\
    \frac{d^2E_j}{d\xi_j^2}&=\sum_{n\neq 0}\frac{|\bra{\chi^j_n}(1-\mathcal{P})V^{(3b)}(1)(1-\mathcal{P})\ket{\chi_0^j}|^2}{E_0-E_n}
\end{align}
where the index $n$ runs over all eigenvectors of the Hamiltonian, and we explicitly write the three-body projection operator discussed in Sec.~\ref{sec:projection}. The first iteration of this algorithm corresponds to calculations considering only two-body interaction, i.e., with $\xi_0=0$.

When projection is not used, the second-order perturbation theory gives rapid convergence in of the binding energy, since the three-body operator is diagonal in $K,\gamma$ and so these operations have a time complexity of $\mathcal{O}(n_{\gamma})$ where $n_\gamma$ is the number of $K,\gamma$ partial waves. The projection operator makes the bra-ket evaluations have a complexity of  $\mathcal{O}(n_{\gamma}^2)$. In this case it is more efficient to use first-order perturbation theory only to update $\xi$, which generally takes 4-6 iterations to converge. 

\section{General B(E$\lambda$) expression for three-body problems} 
\label{sec:multipole}
 
We begin with the general expression \cite{Baye2009}
\begin{align}
    &\frac{dB(E\lambda)}{dE}=\nonumber \\
    &\quad(\hat{J}_0)^{-2}\sum_{S m_SM_0\mu}\int d\ve k_x\, d\ve k_y \, \delta\left(E-\frac{\hbar^2}{2m_N}(k_x^2+k_y^2)\right)\nonumber \\
    &\quad \left|\bra{\Psi^{(-)}_{\ve k_x,\ve k_y,S m_S}(E,\ve x,\ve y)}\mathcal{M}^{(E\lambda)}_\mu\ket{\Psi^{J_0M_0\pi_0}(\ve x,\ve y)}\right|^2,\label{eqn:Belstrength}
\end{align}
where we use the notation $\hat{J}_0=\sqrt{2J_0+1}$, the coordinates $(\ve x,\ve y)$ are shown in Fig.~\ref{fig:TandYbasisdiagram}, and $\mathbf{k_x,k_y}$ are their conjugate wavenumbers, the time-reversed stationary scattering state is defined as
\begin{align}
    &\ket{\Psi^{(-)}_{\ve k_x,\ve k_y,Sm_S}}=  (2\pi)^{-3}\rho^{-5/2}\sum_{JM,K'l_x'l_y'L'M_L'}C_{L'M_L'Sm_S}^{JM}\nonumber \\
    &\qquad \quad \mathcal{Y}_{l_x'l_y'K'}^{L'M_L'*}(\Omega_{5k})\sum_{K\gamma}(-1)^K\ket{\Psi^{J\pi}_{K\gamma(K'\gamma')}}
\end{align}
and  $C$ denotes a Clebsch-Gordan coefficient.

The electric operators \cite{Descouvemont2003} can be written as
\begin{align}
    &\mathcal{M}^{(E\lambda)\mu}(x,y)=\tilde{Z}_y^{(\lambda)} y^\lambda Y_{\lambda}^\mu(\Omega_y)+\tilde{Z}_x^{(\lambda)} x^\lambda Y_{\lambda}^\mu(\Omega_x)\nonumber \\
    &\quad +\sum_{k>0}^{\lambda-1}\alpha_{\lambda k} \tilde{Z}_{xy}^{(\lambda k)}[Y_{k}(\Omega_y),Y_{(\lambda-k)}(\Omega_x)]_{\lambda\mu}y^kx^{\lambda-k},
\end{align}
with
\begin{align}
\alpha_{\lambda k}&=\left(\frac{4\pi(2\lambda+1)!}{(2k+1)!(2\lambda-2k+1)!}\right)^{1/2},\label{eqn:norm}\\
    \tilde{Z}_x^{(\lambda)}&=\frac{e}{\mu_{12}^{\lambda/2}}\left[Z_{2}\left(\frac{-A_1}{A_{12}}\right)^\lambda+Z_1\left(\frac{A_{2}}{A_{12}}\right)^\lambda\right],\\
    \tilde{Z}_y^{(\lambda)}&=\frac{e}{\mu_{(12)3}^{\lambda/2}}\left[Z_{12}\left(\frac{-A_3}{A}\right)^\lambda+Z_3\left(\frac{A_{12}}{A}\right)^\lambda\right],\\
    \tilde{Z}_{xy}^{(\lambda k)}&=\frac{e}{\mu_{12}^{k/2}\mu_{(12)3}^{(\lambda-k)/2}}\left(\frac{-A_3}{A}\right)^k\nonumber \\
    &\qquad \left[Z_2\left(\frac{-A_1}{A_{12}}\right)^{\lambda-k}+Z_1\left(\frac{A_2}{A_{12}}\right)^{\lambda-k}\right].\label{eqn:Z}
\end{align}

Expanding the scattering solutions in their partial waves \cref{eq10}
and doing the same for the bound state $\ket{\Psi^{J_0M_0\pi_0}}$ gives a complete expression 
\begin{widetext}

\begin{align}
&\frac{dB(E\lambda)}{dE}=\frac{(\hat{\lambda})^2m_N^3E2}{(2\pi)^8\hbar^6}\sum_{J}(\hat{J})^2\sum_{K'\gamma'}\left|\sum_{K\gamma;K_0\gamma_0}\delta_{SS_0}(-1)^{L+S+K}\right.\sixj{L_0}{J}{S}{\lambda}{J_0}{L}\hat{L}_0 \,\hat{L}\,\hat{l}_{x0}\,\hat{l}_{y_0}
\int d\rho \,\rho^\lambda \chi^{J\pi}_{K\gamma(K'\gamma')}(\rho)\chi^{J_0\pi_0}_{K_0\gamma_0}(\rho)\nonumber \\
&\qquad\sum_{k=0}^\lambda \alpha_{\lambda k}\tilde{Z}^{\lambda k}\int d\alpha \, \sin^{(2+k)}\alpha\cos^{(2+\lambda-k)}\alpha \phi_{l_xl_y}^K(\alpha)\phi_{l_{x0}l_{y0}}^{K_0}(\alpha) \left.\begin{Bmatrix}
        \lambda-k&k&\lambda\\
        l_x&l_y&L\\
        l_{x0}&l_{y0}&L_0
    \end{Bmatrix}
  \widehat{(\lambda-k)}\, \hat{k}\,C_{l_{x0}0(\lambda- k)0}^{l_x0}C_{l_{y0}0k0}^{l_y0}\right|^2\label{eqn:final}
\end{align}
   
\end{widetext}

\section{Additional convergence tests} \label{AppConvergence}
\begin{figure}[h!]
    \centering
    \includegraphics[width=\linewidth]{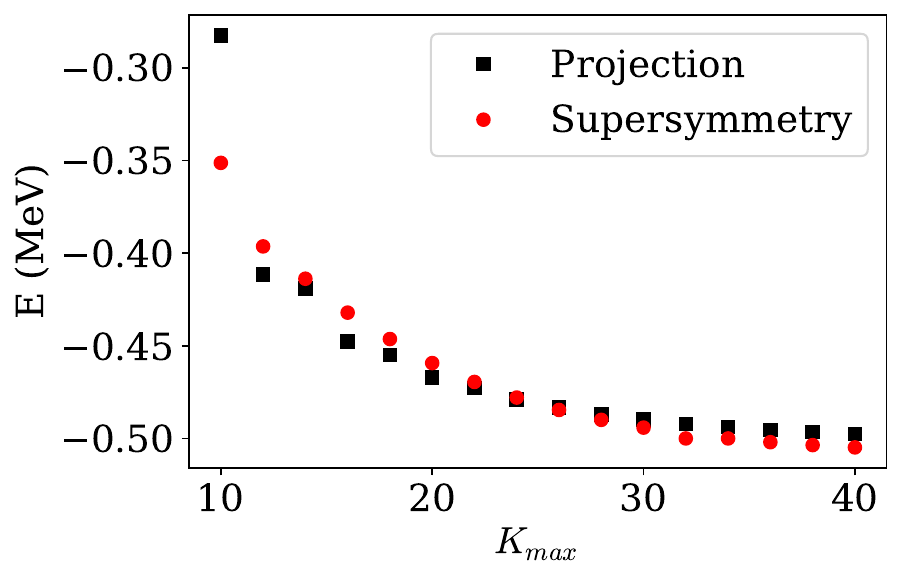}
    \caption{Binding energy of bound states calculated considering the projection (black squares) and supersymmetry (red circles) to remove Pauli-forbidden states and obtained with different $K_{max}$ cutoffs.}
    \label{fig:convergencebound}
\end{figure}

In this appendix, we provide additional information about the convergence of our calculations. To show the convergence of the bound states with the main cutoff $K_{max}$, we do not vary the three-body force, it is fixed to produce a binding energy of 0.5~MeV at $K_{max}=40$ (see ~Sec.\ref{sec:potentials}).  The binding energy obtained with both the projection (black squares) and supersymmetry (red circles) is shown for various $K_{max}$ in Fig.~\ref{fig:convergencebound}.  Interestingly, both methods exhibit similar convergence pattern for the bound state energy for both methods. A convergence to within 100~keV and 10~keV is achieved  respectively by $K_{max}=30$ \and $K_{max}=40$.

\begin{figure}[h!]
    \centering
    \includegraphics[width=\linewidth]{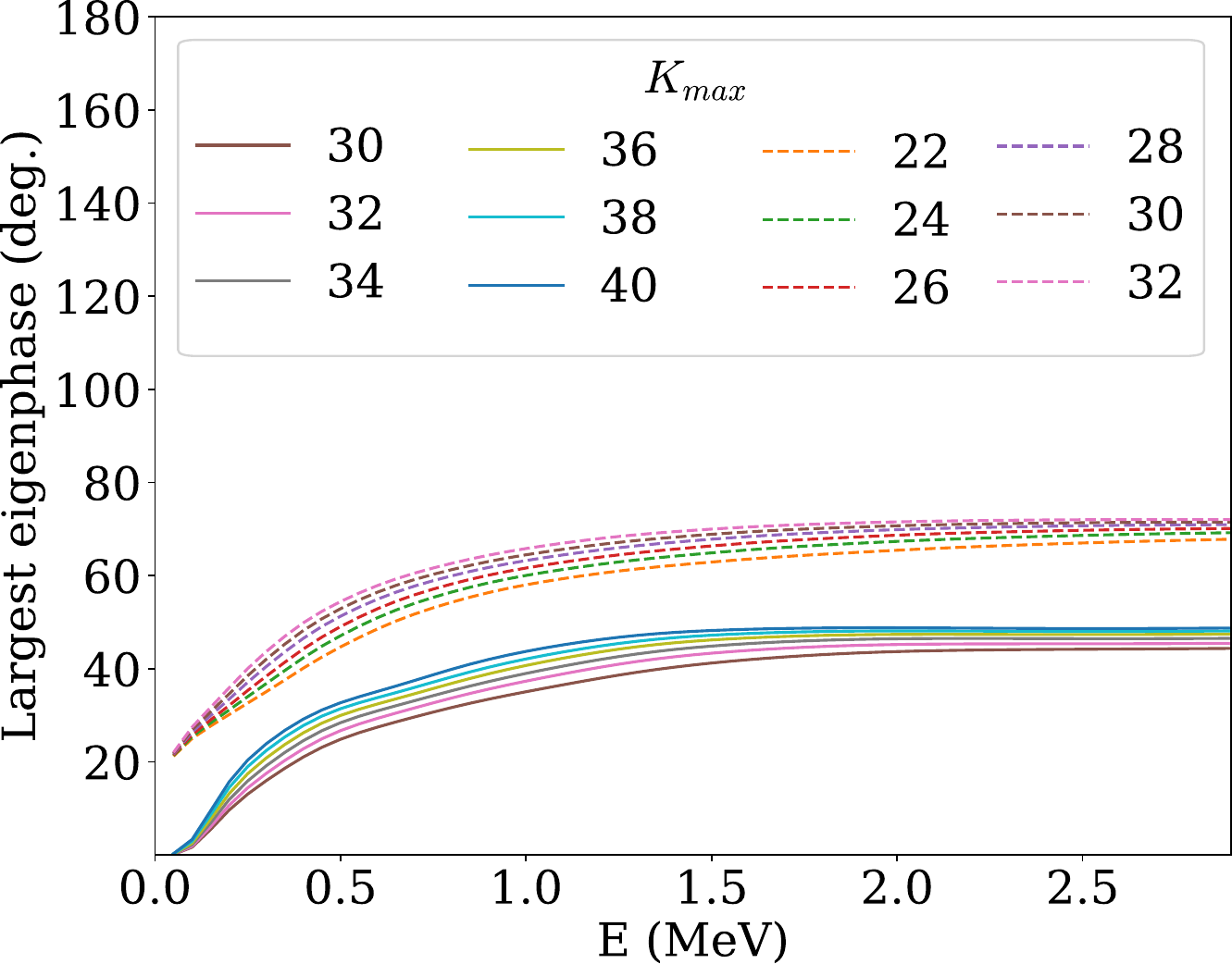}
    \caption{Largest eigenphase $\text{max}(\delta_n)$ \cref{eq:eigenphase} of scattering wave with $J^\pi=1^-$ for each cutoff $K_{max}$, calculated with projection (solid lines) and supersymmetry (dashed lines). There is no three-body force in this channel. To aid the reader, the phaseshifts for supersymmetry are offset by 20$\deg$ from the origin.}
    \label{fig:phaseshift}
\end{figure}

Similarly, the convergence of the scattering states has to be verified. Figure~\ref{fig:phaseshift} shows the largest eigenphase  $\delta_n$~\cref{eq:eigenphase} obtained for different $K_{max}$ and with both the projection (solid lines) and supersymmetry (dashed lines) to remove Pauli-forbidden states. 
 We find that  two approaches lead to similar phaseshifts 
 but the supersymmetry method exhibit a faster convergence than the projection methods, i.e., they converge respectively at $K_{max}=32$ and $K_{max}=40$. Its faster convergence combined with its simpler implementation made the supersymmetry a tool of choice for many few-body models~\cite{Tursunov2016,Lovell2017,Descouvemont2020,Casal2019,Singh2024}. Nevertheless, as shown in Fig. \ref{fig:convergenceBE1} and discussed in the text, they lead to different dipole strengths.

The convergence of the three-body calculations also depends on the parameters of the hyperradial mesh: $\rho_{max}$ and $N$. Since scattering states having a larger amplitude at large $\rho$, they are much more affected by these hyperradial mesh parameters. 
\begin{figure}[h!]
    \centering
    \includegraphics[width=\columnwidth]{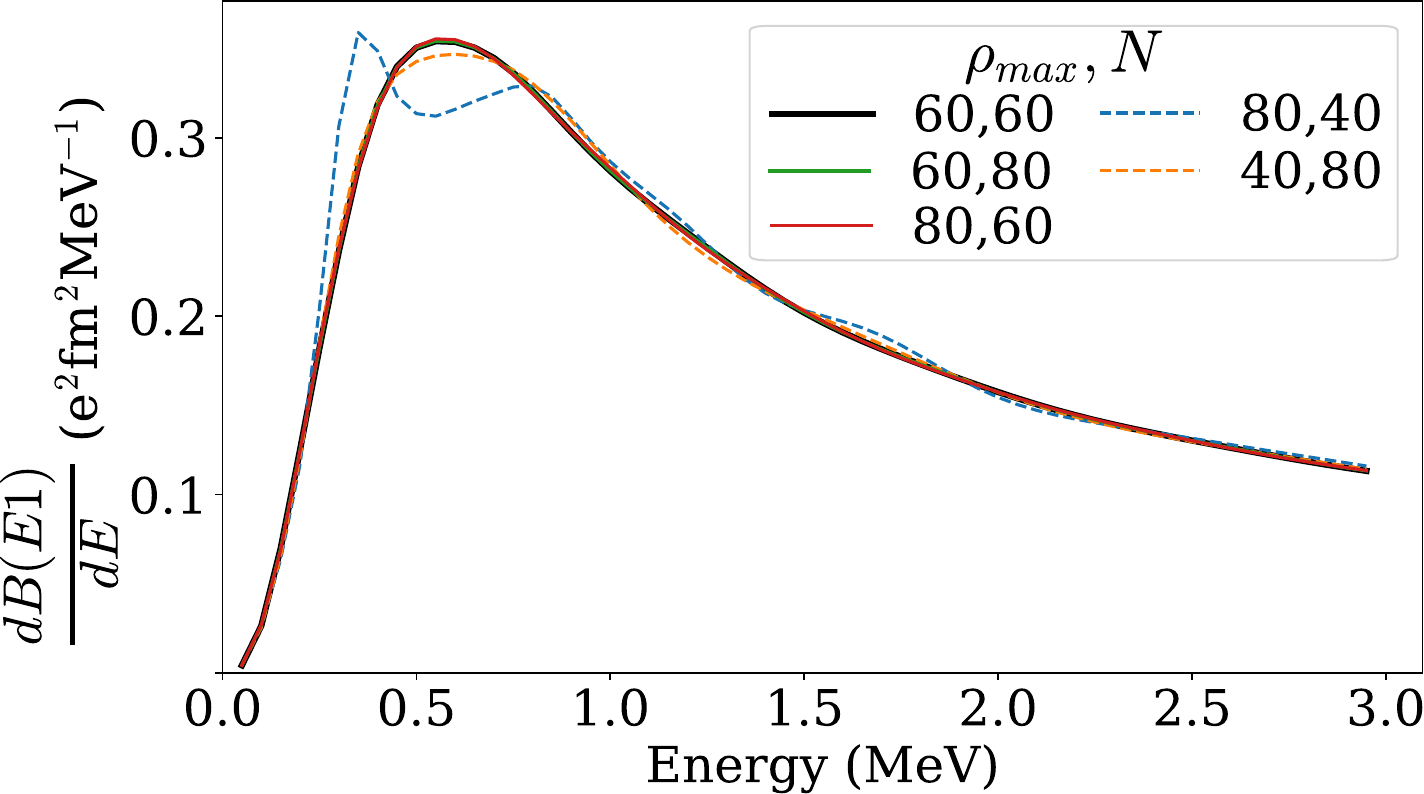}
    \caption{$B(E1)$ strength functions calculated with different cutoffs in $\rho_{max}$ and $N$. All calculations are with $K_{max}=30$ (and have no cutoff on $n$ or $l$ defined in Sec.\ref{Sec:Speedup}).}
    \label{fig:convergencerhoN}
\end{figure}
Figure \ref{fig:convergencerhoN} demonstrates that using too few points in the Lagrange mesh ($N<\rho_{max}/1$~fm) results in unphysical oscillations in the dipole strength. It also shows that the contribution at large hyperradii  are negligible beyond 60 fm. From Fig.~\ref{fig:convergencerhoN}, we establish that using $\rho_{max}=60$~fm with $N=60$ leads to converged results for the dipole strength.

\section{Parameters of two-body force}
Table~\ref{tab:parameters} contains the potential parameters used in this work to define the core-neutron interaction, which adjustment is detailed in Sec.~\ref{sec:potentials}. 

\begin{table}[h!]
    \centering
    \begin{tabular}{c|cccc}
    $l$&$R_l$ (fm)&$a_l$ (fm)&$V^0_l$ (MeV)&$V^{so}_l$ (MeV)\\ \hline 
    0 & 3.06& 0.65&35.1 & 17.0\\
    1 & 3.06& 0.65&41.1 & 17.0\\
    2 & 3.06& 0.65&47.2 & 17.0\\
    $>2$ & 3.06& 0.65&41.1 & 17.0\\
    \end{tabular}
    \caption{Two-body core-neutron potential parameters for all partial waves.}
    \label{tab:parameters}
\end{table}
\clearpage

\section{Occupations}
Table~\ref{tab:occupations} contains the occupations of the four largest partial waves in each basis (T and Y) of the bound states calculated and displayed in Fig. \ref{fig:occupations}. Note that the amplitudes are summed over $K$.

\begin{table}[h!]
    \centering
    \begin{tabular}{c|c|c}
      &Projection  &  Supersymmetry\\ \hline \hline
      ($l_x,l_y$)& &\\ 
      (0,0)  &  0.646 & 0.952 \\
       (1,1)  & 0.011  &  0.027\\
      (2,2) &   0.309& 0.019 \\
      (4,4) & 0.029 & 0.001\\
      \hline 
      ($l_a,l_b$) &&\\
      (0,0)& 0.954& 0.878\\
      (1,1)&0.016 & 0.036\\
      (2,2)&0.022 & 0.072\\
      (3,3)&0.006 & 0.010\\
    \end{tabular}
    \caption{Amplitudes of the four strongest partial waves in the $0^+$ ground state at $S_{2n}=0.5$~MeV calculated with projection and supersymmetry. $l_x,l_y$ angular momenta are in the T-basis and $l_a,l_b$ angular momenta are in the Y-basis.}
    \label{tab:occupations}
\end{table}

\end{document}